\documentclass[journal]{IEEEtran}
\ifCLASSINFOpdf
\usepackage[pdftex]{graphicx}
\graphicspath{{../pdf/}{../jpeg/}}
\DeclareGraphicsExtensions{.pdf,.jpeg,.png}
\else
\usepackage[dvips]{graphicx}
\graphicspath{{../eps/}}
\DeclareGraphicsExtensions{.eps}
\fi
\usepackage{graphicx}
\usepackage{graphics}
\usepackage{epsfig}
\usepackage{epstopdf}
\usepackage{stfloats}
\usepackage[cmex10]{amsmath}
\usepackage{algorithmic}
\usepackage{algorithmic}
\usepackage[ruled]{algorithm2e}
\SetKwRepeat{Do}{do}{while}
\usepackage{array}
\usepackage{mdwmath}
\usepackage{dsfont}
\usepackage{mdwtab}
\usepackage{graphicx}
\usepackage{subfigure}
\usepackage{color}
\usepackage{amsfonts,amssymb}
\usepackage{hyperref} %
\usepackage{lineno} 
\usepackage{multirow}
\usepackage{diagbox}
\usepackage{caption}
\usepackage{bbding}
\usepackage{amsfonts,amsthm,array} 
\usepackage[ruled]{algorithm2e}
\usepackage{makecell}
\usepackage{threeparttable}
\usepackage{bm} 

\usepackage{cuted}

\usepackage{makecell}

\usepackage{bbding}
\usepackage{newtxmath}  

\begin{document}
	

\title{Two Birds With One Stone: Beamforming Design for Target Sensing and Proactive Eavesdropping\thanks{Manuscript received.}}

\author{Qian~Dan, 
	Hongjiang~Lei,
	Ki-Hong~Park, 
	Gaofeng~Pan, 
	and~Mohamed-Slim~Alouini 
	\thanks{This work was supported by the National Key Research and Development Program of China under Grant 2024YFC3306801 and Natural Science Foundation of Chongqing, China under Grant CSTB2025NSCQ-LZX0053. (Corresponding author: \textit{Hongjiang~Lei}).}
	\thanks{Qian~Dan is with the School of Communications and Information Engineering, Chongqing University of Posts and Telecommunications, Chongqing 400065, China, and also with the School of Intelligent Medicine and Information Engineering, Jiangxi University of Chinese Medicine, Nanchang 330004, China (e-mail: {danqian@jxutcm.edu.cn}).}
	\thanks{Hongjiang~Lei is with the School of Communications and Information Engineering, Chongqing University of Posts and Telecommunications, Chongqing 400065, China
		(e-mail: leihj@cqupt.edu.cn).}	
	\thanks{Gaofeng~Pan is with the School of Cyberspace Science and Technology, Beijing Institute of Technology, Beijing 100081, China (e-mail: gfpan@bit.edu.cn).}
	\thanks{Ki-Hong~Park and Mohamed-Slim~Alouini are with CEMSE Division, King Abdullah University of Science and Technology (KAUST), Thuwal 23955-6900, Saudi Arabia (e-mail: kihong.park@kaust.edu.sa, slim.alouini@kaust.edu.sa).}
}

\maketitle

\begin{abstract}
	
This work studies the beamforming design in the joint proactive eavesdropping (PE) and target sensing (TS) systems.
The base station (BS) wiretaps the information transmitted by the illegal transmitter and sends the waveform for TS.
The shared waveform also serves as artificial noise to interfere with the illegal receiver, thereby achieving successful PE.
We firstly optimize the transmitting beampattern of the BS only to maximize the eavesdropping rate or only to minimize the Cram{\'{e}}r-Rao bound, respectively.
Then, the joint design of PE and TS is investigated by formulating the PE-centric, the TS-centric, and the normalized weighted optimization problems.
The formulated problems are solved by the semi-definite relaxation technique and the sequential rank-one constraint relaxation method to address the complexity of the original problem.
Furthermore, the scenario in which the quality of the eavesdropping channel is stronger than that of the illegal channel is considered.
Numerical simulation demonstrates that the proposed algorithm can effectively realize PE and TS simultaneously.	
	
\end{abstract}


\section{Introduction}
\label{sec:introduction}	

\subsection{Background and Related Works}
\label{sec:Background}	
	
Integrated sensing and communication (ISAC) technology can share spectrum, hardware platforms, and even baseband waveforms and signal processing between communication and sensing to improve the spectral efficiency, energy efficiency (EE), and hardware efficiency of the system for integration gain. Further, the mutual assistance and mutual gain of the two functions can also be used to improve each other's performance and obtain coordination gain \cite{LiuF2022JSAC}-\cite{NiW2025WC}.

Recently, the performance for various ISAC systems was investigated to achieve concurrent sensing and communication, which were classified into the radar-centric systems \cite{LiuF2022TSP}-\cite{ZhaoZ2024CL}, the communication-centric systems \cite{LiuT2022JCIN}-\cite{XiaF2025TWC}, and the joint-design systems \cite{HuaH2024TWC}-\cite{WuG2024IOTEE}.
The authors in \cite{LiuF2022TSP} investigated beamforming designs for the ISAC systems with multiple communication users for scenarios with point or extended targets. In particular, the Cram{\'{e}}r-Rao bound (CRB) of the azimuth angle or the response matrix of the target was minimized by designing the dual-functional beamforming matrix.
In \cite{WenC2023TSP}, a new angular waveform similarity metric was proposed and combined with the integrated main-lobe-to-sidelobe ratio of the transmit beampattern to evaluate the waveform ambiguity properties.
The complementary waveform and linear precoder matrix were jointly designed to minimize the weighted sum of the integrated main-lobe-to-sidelobe ratio while considering a predefined signal-to-interference-plus-noise ratio (SINR), power, and peak-to-average-power ratio constraints.
In \cite{RenZ2024TWC}, the authors investigated the performance of the ISAC systems with multiple targets and communication users.
Both the scenarios without and with prior target knowledge at the base station (BS) were considered, and the multi-target response matrix and the reflection coefficients/angles were estimated, respectively.
The CRB of multi-target estimation was minimized while satisfying the constraints of a minimum multicast communication rate and a maximum transmit power.
The authors in \cite{ZhaoZ2024CL} proposed a new robust beamforming scheme for ISAC systems with different point targets.
The maximum CRB of the estimated direction of arrival was minimized by designing the beamforming matrix while ensuring the quality-of-service (QoS) for all the communication users. 
The authors in \cite{LiX2023TWC} proposed an integrated sensing, communication, and computation over-the-air framework to enable simultaneous communication, target sensing (TS), and over-the-air computation.
The shared and separated schemes were proposed, and the accuracy of over-the-air computation was optimized by jointly designing the data transmission beamformer and the aggregation beamformer or the data beamformer and the radar sensing beamformer.
In \cite{LiuT2022JCIN}, the authors proposed a new waveform design based on constructive interference for the ISAC system with multiple communication users.
Their results show that the proposed scheme can reduce the transmit power for the given constraint on estimation accuracy or increase the communication signal-to-noise ratio (SNR) with a given power budget. 
The authors in \cite{LiuRang2023TWC}  investigated the SNR-constrained and CRB-constrained joint beamforming and reflection design problems for the reconfigurable intelligent surface (RIS)-aided ISAC systems.
The beamforming matrix and RIS reflection matrix were jointly optimized to maximize the sum of communication users' communication rate while considering the SNR/CRB, the BS's transmission power, and the RIS's unit-modulus constraints. 
In \cite{XiaF2025TWC}, the authors proposed a symbiotic sensing and communication framework composed of base stations (BS) and passive sensing nodes. It considers both fully digital arrays and hybrid analog-digital (HAD) arrays, and maximized the achievable sum rate through beamforming design under the constraint of the Cramér-Rao Lower Bound (CRLB) for two-dimensional angle-of-arrival estimation.
In \cite{HuaH2024TWC}, the angle and the reflection coefficient of the target and the CRB of the angle estimation were utilized as the performance metric of the point TS scenario.
The authors adopted three performance metrics, the trace, the maximum eigenvalue, and the determinant of CRB (DCRB), respectively, to estimate the complete target response matrix with the extended TS scenario.
Due to the possibility of mutual interference between the multiplexed signals, there is a trade-off between TS and communication when multi-beam transmitted signals were reused for both sensing and communications.
The EE of the ISAC with an extended target and a multi-antenna communication user was investigated in \cite{WuG2024IOTEE}. 
The energy consumption at the transmitter with the on-off scheme was minimized by jointly designing the transmit covariance matrix and the time allocation while taking the QoS for the communication user and the CRB constraints for the target into account.  	
In \cite{HeZ2023JSAC}, two optimization problems were formulated for the full-duplex ISAC systems.
The power consumption was minimized and the sum rate was maximized by jointly optimizing the downlink dual-functional transmit signal, the uplink receive beamformers at the BS, and the transmit power at the uplink users. 
In \cite{GuoY2024TWC}, the joint active and passive beamforming design problem for the RIS-assisted full-duplex ISAC systems with multiple communication users was considered.
The sum rate was maximized by jointly optimizing the transmit beamformer, the linear postprocessing filters, the users’ uplink transmission power, and the reflected phase shift of RIS. 	
{
	Ref. \cite{DongY2023CL} investigated the joint receiver design for multiple-input multiple-output ISAC, proposing two design strategies for simultaneous symbol detection and target estimation for the radar signals received by the receiver and the signals from the uplink communication users.
	In \cite{OuYangC2023WCL}, the authors analyzed the impact of the successive interference cancellation order on the perceived rate and communication rate in the uplink of a non-orthogonal multiple access ISAC system. 		
}	
	
Like other wireless communication systems, ISAC systems also face security issues, and the confident signals are accessible to be wiretapped \cite{WeiZ2022MagPLS}.
Since the transmit signals were reused for both sensing and communication, the security issue becomes imperative, especially when the target is a potential eavesdropper \cite{SuN2021TWC}-\cite{SuN2024TWC}. 
For the communication-centered works, the secrecy performance was optimized while taking the radar performance and other system constraints into account \cite{SuN2021TWC}-\cite{WeiW2024TWC}. 
In \cite{SuN2021TWC},
the radar target was assumed to be a potential eavesdropper, and the SNR at the potential eavesdropper was minimized by designing the radar beam pattern while ensuring the SINR requirement at legitimate users (LUs).
Both the scenarios with perfect/imperfect target angle and the channel state information (CSI) were considered, respectively.
In 	\cite{SuN2022TWC}, destructive interference was utilized to suppress the potential eavesdropper in the ISAC systems, and the SINR of the radar signals at the BS was maximized by jointly designing the transmit and receive beamforming matrices.
Assuming the estimation error of the azimuth angle of the potential eavesdropper obeys the Gaussian distribution, the authors in \cite{JiaH2023TVT} proposed a new CSI error model.
Both the scenarios with the perfect eavesdropping CSI/a single LU and with imperfect eavesdropping CSI/multiple LUs were considered.
The sum secrecy rate of the considered ISAC system was maximized by optimizing a beamforming vector while considering the CRB and QoS constraints.
In \cite{WeiW2024TWC}, the authors investigated the secure transmission of  simultaneously transmitting and reflecting (STAR)-RIS-assisted ISAC systems. The transmitted beaming, artificial jamming, and passive transmitted and reflected beam formation for STAR-RIS were jointly designed to maximize the sum secrecy rate while taking the target minimum beam pattern gain constraint into account.

For the sensing-centered works, the weighted sum of beampattern matching errors and cross-correlation patterns was utilized as the optimization objective to meet the practical requirements of sensing performance in different scenarios in the target tracking stage \cite{DongF2023TGCN}-\cite{SuN2024TWC}.
Specifically, the scenarios with perfect and imperfect CSI for the ISAC with multiple communication users and potential eavesdropper were considered in \cite{DongF2023TGCN}. 
The weighted sum of the mean square error (MSE) between the designed and desired beampatterns and the mean-squared cross-correlation pattern was minimized by jointly optimizing the covariance matrix of the transmitted signals.
The authors of Ref. \cite{RenZ2023TCOM} investigated the joint secure transmit beamforming designs for ISAC systems with a single LU and multiple potential eavesdroppers.
The weighted sum of beampattern matching MSE and cross-correlation patterns was minimized by jointly optimizing the information and the sensing covariance matrices. The scenarios with bounded CSI errors and Gaussian CSI errors were considered, and the worst-case secrecy rate and the secrecy outage constraints were considered, respectively.
In \cite{SuN2024TWC}, the authors proposed a novel sensing-aided secure scheme for the ISAC systems to obtain the trade-off between secrecy communication and radar sensing performance.
Firstly, the potential eavesdroppers' direction was estimated using the combined capon and approximate maximum likelihood technique.
Then, the dual-functional beamforming and the artificial noise (AN) matrix were jointly optimized to maximize the normalized weighted sum of the secrecy rate and estimated CRB under the estimated angle errors and the system power budget constraints.

In ISAC systems, the coexisting strong radar signals can be utilized as inherent jamming signals to fight the external eavesdroppers and improve the transmission security \cite{WeiZ2022MagPLS}.
The authors in \cite{ChuJ2023TVT} investigated the joint designs of secure transmit beamforming for ISAC systems with multiple eavesdroppers.
For the scenario with the eavesdropping CSI, the maximum eavesdropping SINR of multiple eavesdroppers was minimized by jointly designing the transmit beamformers of communication and radar systems while satisfying the QoS of LUs', the SINR constraint of radar target echo, and the transmit power budget.
For the scenario without the eavesdropping CSI, an AN-aided transmit beamforming scheme was proposed. The total transmit power of the BS and radar was minimized by jointly designing the transmit beamformer and the covariance matrices of the AN vector at the BS and radar under the SINR constraints of LUs' communication and radar target echoes.
In \cite{ZhangJ2024TWC}, an unmanned aerial vehicle (UAV) was utilized as the RIS-aided ISAC base station (BS) and a secure transmission scheme was proposed to maximize the average achievable rate or the EE by jointly designing the transmit power allocation, the scheduling of users and targets, the phase shifts at RIS, as well as the trajectory and velocity of the UAV. 
In \cite{DanQ2025TCCN}, the authors proposed a physical layer security scheme for the rate splitting multiple access system under the constraint of sensing CRB. For the two scenarios of the eavesdropper's channel being perfect and the channel being unknown, under the premise of taking into account user fairness, the maximum energy efficiency is achieved by jointly optimizing the beam design.   	
Ref. \cite{GangY2025DCN} investigated a UAV-assisted full-duplex ISAC system. The average sum rate was maximized by jointly optimizing communication scheduling, beamforming, and UAV trajectory subject to the sensing SINR constraint.

Proactive eavesdropping (PE), also called wireless surveillance, is an effective method of supervising unauthorized wireless monitoring in the physical layer. To be specific, legitimate eavesdroppers wiretap the illegitimate wireless communication between transmitters and receivers to maintain public safety. 
Compared with traditional physical layer security (PLS), the main differences are summarized as follows \cite{DanQ2024IoT}. 
1) \textbf{Application scenarios}: In PLS scenarios, the communication between transmitters and receivers is assumed to be rightful, the eavesdroppers are illegitimate, and various schemes are designed to prevent information leakage to the malicious or potential eavesdroppers \cite{SuN2021TWC}-\cite{ZhangJ2024TWC}. In PE scenarios, it is presumed that the eavesdroppers are legitimate and approved by government agencies to monitor that the suspicious communication between transmitters and receivers, such as criminals or terrorists who may jeopardize public safety \cite{ZengY2016JSTSP}-\cite{XuJ2017TWC}.	Thus, in PLS scenarios, all the efforts are from the transmitters and/or receivers, while all the works are from the view of eavesdroppers in PE scenarios. 
2) \textbf{Prerequisites}: Two conditions are assumed in PE scenarios. One is that the illegitimate transmitter has the channel state information of the unauthorized link, and there is an adaptive transmission between the suspicious source and the destination. The other is that the eavesdropper has global CSI and can decode unauthorized wireless information with any small error based on perfect self-interference cancellation (SIC) since the eavesdropper fully knows the transmission scheme of the illegal link as well as the encoding method \cite{XuJ2017TWC, HuS2021TWC}. There are no similar assumed conditions in PLS scenarios. 
3) \textbf{Performance metric}: In PLS scenarios, the fundamental metric is the achievable secrecy rate, defined as the (non-negative) difference between the achievable rate of the communication link ($R_D$) and the achievable rate of the eavesdropping link ($R_E$). Based on secrecy rate, there are two main performance metrics: average secrecy rate and secrecy outage probability, which denote the statistical average or time average of secrecy rate and the probability that the secrecy rate falls below a target rate \cite{SuN2021TWC}-\cite{ZhangJ2024TWC}.
In PE scenarios, the relationship between $R_D$ and $R_E$ is crucial. 
When $R_E \ge R_D$, surveillance is successful, and $R_D$ is called the eavesdropping rate (ER). Otherwise, surveillance fails, and the ER is equal to zero \cite{ZengY2016JSTSP}-\cite{XuD2022SJ}.
For the scenarios for content analysis and event-based monitoring, there are two main performance metrics: average ER (AER) and successful eavesdropping probability (SEP), also called as eavesdropping non-interrupt probability \cite{XuJ2017TWC}, \cite{FeiziF2020TCOM}. 	
4) \textbf{Motivation}: The motivation of PLS is to make the secrecy rate as large as possible while the aim in PE is to make $R_D$ as large as possible under the condition $R_E \ge R_D$ \cite{XuJ2017TWC}. In particular, when $R_E < R_D$, the eavesdropper will utilize some scheme (such as transmitting jamming signals) to ensure successful eavesdropping ($R_E \ge R_D$). In the scenarios with  $R_E > R_D$, the eavesdropper will utilize some scheme (such as forwarding signals for $D$) to make $R_D$ as large as possible under the condition $R_E \ge R_D$ \cite{ZengY2016JSTSP}, \cite{GuoD2024TVT}.
In summary, PE is an effective technology in the physical layer to maintain public safety\footnote{
	{In Ref. \cite{YinZ2025TIFS}, one interesting scheme, named anti-intercepting transmission, was proposed. 
		The idea is to use false signals to deceive eavesdroppers and hide the real signals to counter eavesdropping threats. To ensure successful deception, the SINR of the decoy signals at the eavesdropper must be at least as high as that of the true signals. All the schemes used at upper layers, such as encryption, are out of scope for this area.
	}}.

For the PE scenarios, many works propose outstanding schemes that can be divided into two types \cite{ZengY2016JSTSP}. 
One is spoofing relaying-based PE schemes that forward the suspicious signal to the illegitimate receivers, and the other is jamming-based PE schemes that utilize the jamming signals to control the interference at the illegitimate receivers. 
The spoofing relaying-based PE scheme is suitable for scenarios wherein the eavesdropping channel is of higher quality than the illegal channel. For example, the authors in Ref. \cite{ZengY2016JSTSP} proposed a power splitting scheme wherein the signal received at the monitor with multiple antennas was utilized for eavesdropping and spoofing relaying, respectively.   
The AER was maximized by jointly designing the power splitting ratios and the precoding matrix at the monitor.
With the help of an amplify-and-forward spoofing relay, a UAV was utilized to eavesdrop on multiple suspicious terrestrial links in \cite{GuoD2024TVT}. 
The AER was maximized by jointly optimizing the precoder at the relay, the receive combiner at the monitor, and the UAV’s trajectory. 	

For the scenarios wherein the quality of the illegal channel is better than that of the eavesdropping channel, jamming signals are utilized to achieve successful PE \cite{XuJ2017TWC}. In particular, in Ref. \cite{XuD2022SJ}, a full-duplex monitor transmits jamming signals while wiretapping illegitimate messages. The legitimate transmitter cooperated with the monitor to control the interference on the illegitimate receiver. 
The SEP of the multiple carrier interference network was maximized by jointly optimizing the transmit power of the eavesdropper and legitimate receiver, and an algorithm based on the Lagrange duality method was proposed to solve the formulated problem. 
In \cite{FeiziF2020TCOM}, the legitimate transmitter with multiple antennas acted as the proactive eavesdropper and utilized the intended communication signal for a legitimate receiver to affect the illegitimate receiver. The SEP of the system was maximized by jointly designing the transmitting and receiving beamformers at the full-duplex eavesdropper.
In Ref. \cite{ZhangH2020TWC}, two suspicious communication links were simultaneously wiretapped by a particular full-duplex monitor with multiple antennas. 
The achievable ER region for the minimum-MSE receivers was investigated and closed-form expressions of the upper and lower
boundary points were derived. Moreover, they proposed a time-sharing-based jamming scheme to enlarge this region by designing the transmit covariance matrix while taking the power budget into account. 
The authors in \cite{LiB2019TVT} proposed a cooperative PE scheme where the primary and assistant monitors cooperatively wiretapped multiple suspicious links. The eavesdropping EE was maximized by designing jamming power and the scheme between monitors while taking the energy transfer and load balancing constraints into account. 
The authors in Ref. \cite{MoonJ2018TWC} proposed a new cooperative PE scheme wherein multiple AF full-duplex relaying monitors simultaneously received and forwarded the suspicious signals to the center monitor. 
The AER was maximized by jointly optimizing the receive combining vector at the central monitor, the precoders at the relays and the transmit covariance matrix at the jammer.	
Table \ref{table1} outlines the typical works on ISAC.

\begin{table*}
	\centering
	\caption{ \textit{Comparison of Related Works on ISAC}.}	
	\label{table1}
	\begin{threeparttable}
		\resizebox{0.8\textwidth}{!}{
			\begin{tabular}{c|c|c|c|c|c|c}
				\Xhline{1.2pt}
				\textbf{Reference} & \textbf{Communication metric} & \textbf{\makecell[c]{Sensing metric}} & \textbf{\makecell[c]{Angle\\ error}} & \textbf{\makecell[c]{Optimization\\ objectives}}& \textbf{Main parameters} & \textbf{\makecell[c]{System}}  \\
				\hline
				
				\hline
				\cite{SuN2021TWC}  &\makecell[c]{ Eavesdropper SNR  \\and legitimate user SINR }& \makecell[c]{ beampattern matching errors/\\beampattern gains }       & \makecell[c]{\checkmark} &     Received SNR of targets  &   Waveform covariance matrix & \makecell[c]{Multiple users, \\a target\ potential Eve}  \\

				\hline
				\cite{SuN2022TWC}  & SNR of user   & Radar echo SINR        &  &   SINR of radar echo    &  \makecell[c]{Transmit waveform, \\receive beamformer}  &  \makecell[c]{Multiple users,\\ a point-like target}\\
				
				\hline
				\cite{JiaH2023TVT}  & Secrecy rate   & CRB        & \makecell[c]{\checkmark} &    Worst secrecy rate    &    Beamforming vector  & \makecell[c]{Multiple users,\\a single target/Eve} \\	
				
				\hline
				\cite{WeiW2024TWC}  & Sum secrecy rate   & Beampattern gain         & &  Sum secrecy rate     &   \makecell[c]{ Transmit beamformer, \\artificial jamming vector, 
					\\STAR-RIS's vectors}  &\makecell[c]{  Multiple users,\\ a single target }\\

				\hline
				\cite{DongF2023TGCN}  & SINR of users and Eve  & \makecell[c]{Beampattern matching MSE\\ and cross-correlation pattern   }       &  &  \makecell[c]{Weighted sum
					of \\beampattern matching errors\\ and cross-correlation patterns  } &   \makecell[c]{Transmit beamformer}  & \makecell[c]{Multiple users,\\ multiple targets }\\

				\hline
				\cite{RenZ2023TCOM}  &Achievable secrecy rate   & \makecell[c]{Beampattern matching MSE \\and cross-correlation pattern   }       & & \makecell[c]{ Weighted sum
					of\\ beampattern matching errors \\and cross-correlation patterns }  &     \makecell[c]{Transmit covariance matrix and \\ the sensing/AN covariance matrix }  &\makecell[c]{ Single  user,\\ multiple targets }\\

				\hline
				\cite{SuN2024TWC}  &Secrecy rate   & CRB        & \checkmark &  \makecell[c]{ Weighted sum
					of CRB\\ and  secrecy rate }    & Transmit covariance matrix   &  \makecell[c]{Multiple users,\\multiple targets/Eves} \\
				
				\hline
				\cite{ChuJ2023TVT}  & \makecell[c]{Communication SINR \\and eavesdropping SINR }  & Radar output SINR        &  &\makecell[c]{  Eavesdropping SINR/\\ transmit power  }   &    \makecell[c]{BS transmit beamformer, \\radar transmit beamformer,	\\ radar receive filter }  &\makecell[c]{ Multiple eavesdroppers, \\
					multiple users }\\

				\hline
				\cite{ZhangJ2024TWC}  & Average achievable rate   & Received power        &  &   Average achievable rate   &  \makecell[c]{ Transmit power allocation,\\ users and target scheduling,\\ IRS's phase shifts,\\
					trajectory and velocity of the UAV }& \makecell[c]{ Multiple users,\\  multiple targets} \\				
				\hline
				Our work   & ER &  CRB  & \checkmark  & SNR, CRB &\makecell[c]{ Waveform covariance matrix  } & \makecell[c]{ An illegal user,\\ multiple targets}   \\
				\Xhline{1.2pt}
		\end{tabular}}
	\end{threeparttable}
\end{table*}

\subsection{Motivation and Contributions}
\label{sec:Motivation}

As stated in the previous subsection, in PE scenarios wherein the quality of the eavesdropping link is weaker than that of the illegitimate communication link ($R_E < R_D$), the eavesdropper will utilize some scheme (such as transmitting AN) to ensure successful eavesdropping ($R_E \ge R_D$). 
Moreover, in ISAC systems, the strong radar signals have been utilized as inherent jamming signals to fight the external
eavesdroppers and improve the secrecy performance of the legitimate. 
The above works implement communication and sensing functions within the ISAC system, but the results can not be utilized in PE scenarios because of the different motivations between the PLS and PE. 
Thus, we have the following questions: 
\textit{
	How do we utilize the radar signals in PE scenarios? 
	What is the effect of joint PE and TS? 
}
To answer these questions, we propose two waveform design schemes: one that maximizes the ER and the other that minimizes the DCRB.
Then, a normalized weighted optimization problem is formulated to optimize PE and TS jointly. 
\textit{Relative to realizing the trade-off between PLS and TS in many outstanding works, such as 
	\cite{RenZ2024TWC}, \cite{SuN2021TWC}, \cite{SuN2022TWC},
	\cite{WeiW2024TWC},  \cite{DongF2023TGCN},  \cite{RenZ2023TCOM}, 	
	\cite{SuN2024TWC}, etc., the results in this work demonstrate that the proposed beamforming scheme can catch two birds (PE and TS) simultaneously.}

For clarity, the contributions of this work are listed as follows.
\begin{enumerate}
	\item  We consider the scenario in which a multi-antenna BS actively wiretaps the illegal links while sensing multiple targets.
	The beam transmitted by the BS for TS is also utilized as an AN to interfere with illegal links and enable successful PE.
	We firstly optimize the transmitting beam of the BS only to maximize the ER or only minimize the DCRB, respectively.
	The semi-definite relaxation (SDR) method was employed to solve the PE and TS optimization problems after dropping the rank-one constraint.
	
	\item To realize PE and TS simultaneously, the PE-centric, TS-centric, and the normalized weighted optimization problems are formulated while taking the maximum transmit power constraint into account. 
	All the formulated problems are solved by the SDR technique and the sequential rank-one constraint relaxation (SROCR) method.
	Furthermore, the scenario in which the quality of the eavesdropping channel is stronger than that of the illegal channel is considered.	
	
	\item Relative to realizing the trade-off between PLS and TS in many outstanding works, such as 
	\cite{RenZ2024TWC}, \cite{SuN2021TWC}, \cite{SuN2022TWC},
	\cite{WeiW2024TWC},  \cite{DongF2023TGCN},  \cite{RenZ2023TCOM}, 	
	\cite{SuN2024TWC}, the proposed scheme can effectively realize PE and TS simultaneously. 
	The sensing signals are utilized as AN in these works to suppress the illegitimate receivers, and in theory, the more interference to the illegitimate receivers, the better the secrecy performance. 
	\textit{There is a trade-off between communication and sensing performance in ISAC systems. }
	However, in our work, the sensing signals need to be designed carefully since excessive interference to the illegitimate receivers will result in lower ER. 
	\textit{To the best of the authors' knowledge, no open literature addresses the joint TS and PE optimization problem. More interestingly, the results in this work demonstrated that the proposed scheme can achieve a win-win in the joint TS and PE scenario.}
	
	\item Relative to Refs. \cite{DanQ2024IoT}-\cite{MoonJ2018TWC} wherein the outstanding schemes (such as spoofing relaying-based and jamming-based schemes) were proposed to achieve successful PE \textit{with some additional resources} (such as the relay or transmitting jamming signals), this work utilizes the \textit{existing} sensing signals to achieve successful PE while completing the sensing function. 	
	Thus, more integration gain (especially EE) and coordination gain are obtained. 	
\end{enumerate}

\subsection{Organization}

The remaining organization of this paper is summarized as follows. 
Sect. \ref{sec:SystemModel} introduces the system model of the joint PE and TS communication.
In Sects. \ref{sec:Problem1} and \ref{sec:Problem3}, the PE-centric, TS-centric, and the normalized weighted optimization problems are formulated and solved by the SDR technique and the SROCR method. 
In Sect. \ref{sec:Simulation}, the simulation results of the algorithm are presented and analyzed.
Finally, Sect. \ref{sec:Conclusions} summarizes this work.
TABLE \ref{table2} lists the notations and symbols utilized in this work.

\textit{Notations:} Vectors and matrices are represented in lowercase and uppercase boldface letters, respectively, 
$\mathbf I$ denotes an identity matrix with appropriate dimensions, 
${\left(  \cdot  \right)^{-1}}$ and ${\left(  \cdot  \right)^T}$ represent the inverse and transpose of the matrix, respectively, 
${\left(  \cdot  \right)^*}$  and ${\left(  \cdot  \right)^H}$ represent the conjugate and conjugate transpose of the matrix, respectively, 
${\mathrm{tr}}\left(  \cdot  \right)$ represents the trace of a matrix, 
$\left( \cdot \right)_{\left(i,j\right)}$denotes the $\left(i, j\right)$-th element of the matrix, 
$ \odot $ denotes Hadamard product, 
and 
${\mathop{\mathrm {Re}}\nolimits} \left(  \cdot  \right)$ and $ {\mathop{\mathrm {Im}}\nolimits} \left(  \cdot  \right)$ signify
the real and the imaginary parts of a complex-valued matrix, respectively.			

\begin{table}
	\caption{\textit{List of main symbol Notations.}}
	\begin{center}
		\begin{tabular}{c| c }
			\Xhline{1.2pt}
			\textbf{Notation}   	& \textbf{Description}								\\
			\hline
			$d_{SD}$             & Distance between $S$ and $D$\\
			\hline
			$d_{SE}$              & Distance between $S$ and $E$	\\
			\hline	
			$d_{ED}$         & Distance between $E$ and $D$					\\
			\hline
			$a_r$, $a_t$     & Steering vector of the transmitting/receiving arrays  				\\
			\hline
			$\beta_k$			& \makecell[c]{Normalized coefficients for target reflection \\complex amplitude and path attenuation of target $T_k$	}				\\
			\hline
			${\bm A}_r$, ${\bm A}_t$     & \makecell[c]{Steering matrix for multiple targets angles\\of the transmitting/receiving arrays  }				\\
			\hline
			${\bm \theta}$			& Vector of multiple target angles\\
			\hline
			$\bm \beta$			& \makecell[c]{Normalized coefficients for target reflection complex \\amplitude and path attenuation of multiple target angles	}				\\
			\hline
			$\theta_{T_k}$   &Azimuth of the pre-estimated target $T_k$\\
			\hline
			$\theta_D$&   Azimuth of $D$\\
			\hline
			$\Omega$ & Detection range except the target angle $\theta _{T_k}$\\
			\hline
			$\Omega_1$&  Detection range except the mainlobe\\
			\hline
			$\Omega_2$&  Mainlobe\\
			\hline
			$d$               & Antenna spacing 							\\
			\hline
			$\lambda$ 					& Wavelength			\\ 
			\hline
			$\sigma_D^2, \sigma_E^2$ 					& Receiver noise power			\\ 
			\hline
			$\sigma^2$ & Radar noise power\\
			\hline
			$\beta_0$&    Channel gain at the reference distance\\
			\hline
			$\alpha$&   Path loss exponent\\
			\hline
			$N_t$, $N_r$&     Number of transmitting/receiving antennas \\
			\hline
			$\rho$ &         Weighting factor\\
			\hline
			$\varphi$ &     Scalar associated with the wide main beam fluctuation  \\
			\hline
			$\gamma _s$ &     \makecell[c]{Requirement between the gain of \\ the target direction and the gain of other directions}\\
			\hline
			$\alpha_{\Phi}$&  Threshold of the DCRB\\
			\hline
			$\alpha_{I}$&  Threshold of $D$' angle beampattern gains \\
			\hline
			$\vartheta$ & Convergence accuracy\\
			\hline
			$P_S$, $P_0$ & Transmit power of $S$ and $E$\\	
			\Xhline{1.2pt}
		\end{tabular}
	\end{center}
	\label{table2}
\end{table}

\section{System Model}
\label{sec:SystemModel}

\begin{figure}[t]
	\centering		
	\includegraphics[width = 0.253 \textwidth]{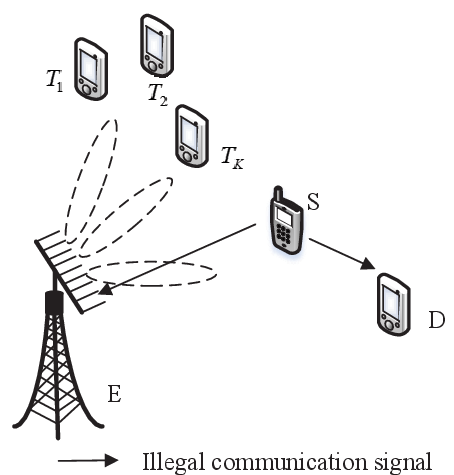}
	\caption{System model consisting of an illegal source $\left( S \right)$, a destination $\left( D \right)$,  $K$ targets $\left( {{T_k},k = 1, \cdots, K} \right)$ and a BS $\left( E \right)$.}
	\label{fig_model}
\end{figure}

As shown in Fig. \ref{fig_model}, we consider a joint TS and PE system which consists of a single-antenna illegal transmission source node $\left( S \right)$, a single-antenna destination node $\left( D \right)$, {$K$ sensing targets $\left( {{T_k},k = 1, \cdots, K} \right)$}, and a BS $\left( E \right)$ equipped with uniform linear array with $N_t$ transmitting and $N_r$ receiving antennas. 
There is an illegal link for adaptive transmission between $S$ and $D$ \cite{XuD2022SJ}, \cite{FeiziF2020TCOM},
and $E$ transmits signals simultaneously to sense { $T_k$} and wiretap the information that $S$ sends to $D$. 	
It is assumed that the location of $S$ and $D$, the CSI of $S$-$D$, and azimuth angle of $T_k$ are available at $E$\footnote{
	Note that these information can be obtained through the preceding detection and this assumption has been utilized in lots of literature focused on ISAC, such as \cite{RenZ2024TWC}, \cite{RenZ2023TCOM}. 
	Explicitly, the location of $S$ can be easily estimated by its transmitted signals. The locations of $D$ and $T_k$ can be obtained with the proposed schemes in some outstanding works on the ISAC systems, such as \cite{RenZ2024TWC} \cite{HuaH2024TWC}, \cite{RenZ2023TCOM}, \cite{SuN2024TWC}.
}.
The signals transmitted by $E$ to sense $T_k$ is also utilized as a controllable AN to achieve successful PE. 
Thus, the beam is designed to improve the accuracy of the estimation as well as the PE performance.  

\subsection{Target Sensing Model}

The transmitted signal, ${{\mathbf x}_E \in \mathbb{C}^{N_t\times 1}}$, at $E$ is expressed as
\begin{align}
	{\mathbf x}_E = {\mathbf{w}} {s},
	\label{xE}
\end{align}
where 
${\mathbf w} \in \mathbb{C}^{N_t\times 1}$ is the transmit beamforming to be designed 
and 
${s}$ denotes the symbols with unit power. 
The echo received by the radar is expressed as\footnote{
	{
		Like \cite{HeZ2023JSAC, ChuJ2023TVT}, since the radar host generally has a strong signal separation capability, and a variety of methods can be used for filtering, $E$ is assumed to perfectly separate the signal (the echo from $S$) from the interference (the echo from $D$). 
		It should be noted that this is called the worst-case scenario for PLS \cite{LeiH2020TCOM} but is the best-case scenario for PE since this assumption overestimates the eavesdropper's decode-ability in many scenarios.
	}
}
\begin{align}	
	{\mathbf{Y}}\left( {\bm{\theta }} \right) = {{\mathbf{A}}_r}\left( {\bm{\theta }} \right){\bm{\beta A}}_t^T\left( {\bm{\theta }} \right){{\mathbf{x}}_E} + {\mathbf{Z}},
	\label{Ytheita}
\end{align}
where 	
${\bm{\theta }} = \left[ {{\theta _{T_1}},...,{\theta _{T_K}}} \right]^T$, 		 
$\theta _{T_k}$ is the angle of the $k$-th target\footnote{
	{
		The parameter estimation consists of rough estimation, waveform design, and fine estimation, which form a closed-loop optimization logic. 
		This work focuses on waveform design, which serves as a ``bridge'' connecting rough and fine estimation. Based on the rough estimation results, a transmission waveform that is insensitive to errors is designed to enhance the recognizability of the target parameters in the received signal.
	}
} 
wherein the first antenna is utilized as the reference point,
$k = 1, \cdots, K$, 
${{\mathbf{A}}_r} = \left[ {{{\mathbf{a}}_r}\left( {{\theta _{T_1}}} \right),...,{{\mathbf{a}}_r}\left( {{\theta _{T_K}}} \right)} \right]$, 
${{\mathbf{A}}_t} = \left[ {{{\mathbf{a}}_t}\left( {{\theta _{T_1}}} \right),...,{{\mathbf{a}}_t}\left( {{\theta _{T_K}}} \right)} \right]$,	 
${{\mathbf a}_t}\left( {{\theta _{T_k}}}  \right) = [1,{e^{j\frac{{2\pi d}}{{\lambda }}\sin \left( {\theta _{T_k}}  \right)}}, ..., {e^{j\frac{{2\pi d}}{{\lambda }}\left( {{N_t}-1} \right)\sin \left( {\theta _{T_k}}  \right)}}]^T$, 
${{\mathbf a}_r}\left( {{\theta _{T_k} }}  \right) = [1,{e^{j\frac{{2\pi d}}{{\lambda }}\sin \left( {\theta _{T_k}}  \right)}},...,{e^{j\frac{{2\pi d}}{{\lambda }}\left( {{N_r}-1} \right)\sin \left( {\theta _{T_k}}  \right)}}]^T$, 
$d$ is the antenna spacing,
$\lambda$ denotes the wavelength, 
${\bm{\beta }} = {\mathrm {diag}}\left( {{{\left[ {{\beta _1},...,{\beta _K}} \right]}^T}} \right)$, 	 
$\beta_k  = {\beta _{R_k}} + j{\beta _{I_k}}$ denotes the normalized parameter which depends on the complex amplitude of the target reflection and path loss of the $S-D$ link \cite{LiuF2022TSP}, \cite{RenZ2024TWC}, \cite{RenZ2023TCOM},  
${\mathbf Z}\sim CN\left( 0,\sigma ^2{\mathbf{I}} \right) $, 
and $\sigma^2$ is noise power. 	
According to (\ref{Ytheita}), ${\mathbf{Y}}\left( {\bm{\theta }} \right) \sim CN\left( {{{\mathbf{A}}_r}\left( \bm \theta  \right)\bm\beta {\mathbf{A}}_t^H\left( \bm \theta  \right){{\mathbf x}_E},\sigma ^2{\mathbf{I}}} \right)$.

{It must be noted that, although the distance among all the nodes, and the pre-estimated azimuth angle and the reflection amplitude are available at $E$, the aim of TS in this work is to improve the estimation performance of azimuth and reflection amplitude information through carefully design of the radar waveform. 
	We define ${\bm \zeta } = \left[ {{{\bm\theta }}, {\bm\beta _R}, {{{\bm\beta }}_I}} \right] $.
}
Since $\sigma ^2{\mathbf{I}}$ is independent of the estimator \cite{RenZ2024TWC, SuN2024TWC, LiJ2008TSP},
therefore, the Fisher information matrix (FIM) for ${\bm \zeta } $ is expressed as
\begin{align}
	{\mathbf{F}}\left( {{\bm \zeta }} \right) = \frac{2}{{{\sigma ^2}}}\left[ {\begin{array}{*{20}{c}}
			{{\mathop{\mathrm {Re}}\nolimits} \left( {{{\mathbf{F}}_{11}}} \right)}&{{\mathop{\mathrm {Re}}\nolimits} \left( {{{\mathbf{F}}_{12}}} \right)}&{-{\mathop{\mathrm {Im}}\nolimits} \left( {{{\mathbf{F}}_{12}}} \right)}\\
			{{{{\mathop{\mathrm {Re}}\nolimits} }^T}\left( {{{\mathbf{F}}_{12}}} \right)}&{{\mathop{\mathrm {Re}}\nolimits} \left( {{{\mathbf{F}}_{22}}} \right)}&{-{\mathop{\mathrm {Im}}\nolimits} \left( {{{\mathbf{F}}_{22}}} \right)}\\
			{-{{{\mathop{\mathrm {Im}}\nolimits} }^T}\left( {{{\mathbf{F}}_{12}}} \right)}&{-{{{\mathop{\mathrm {Im}}\nolimits} }^T}\left( {{{\mathbf{F}}_{22}}} \right)}&{{\mathop{\mathrm {Re}}\nolimits} \left( {{{\mathbf{F}}_{22}}} \right)}
	\end{array}} \right],
	\label{FIM2}
\end{align}
where
${F_{11}} =\left( {{\mathbf{A}}_r^H{{\dot{\mathbf{ A}}}_r}} \right) \odot \left( {{{\bm{\beta}} ^*}\dot{\mathbf{ A}}_t^H{{\mathbf{R}}^*}{{\mathbf{A}}_t}{\bm{\beta}} } \right) + \left( {{\mathbf{A}}_r^H{{\mathbf{A}}_r}} \right) \odot \left( {{{\bm{\beta}} ^*}\dot{\mathbf{ A}}_t^H{{\mathbf{R}}^*}{{\dot{\mathbf{ A}}}_t}{\bm{\beta}} } \right) + \left( {\dot{\mathbf{ A}}_r^H{{\dot{\mathbf{ A}}}_r}} \right) \odot \left( {{{\bm{\beta}} ^*}{\mathbf{A}}_t^H{{\mathbf{R}}^*}{{\mathbf{A}}_t}{\bm{\beta}} } \right) + \left( {\dot{\mathbf{ A}}_r^H{{\mathbf{A}}_r}} \right) \odot \left( {{{\bm{\beta}} ^*}{\mathbf{A}}_t^H{{\mathbf{R}}^*}{{\dot{\mathbf{ A}}}_t}{\bm{\beta}} } \right)$,
$	{F_{12}} = \left( {{\mathbf{A}}_r^H{{\mathbf{A}}_r}} \right) \odot \left( {{{\bm{\beta}} ^*}\dot{\mathbf{ A}}_t^H{{\mathbf{R}}^*}{{\mathbf{A}}_t}} \right) + \left( {\dot{\mathbf{ A}}_r^H{{\mathbf{A}}_r}} \right) \odot \left( {{{\bm{\beta}} ^*}{\mathbf{A}}_t^H{{\mathbf{R}}^*}{{\mathbf{A}}_t}} \right)$,
${F_{22}} = \left( {{\mathbf{A}}_r^H{{\mathbf{A}}_r}} \right) \odot \left( {{\mathbf{A}}_t^H{{\mathbf{R}}^*}{{\mathbf{A}}_t}} \right)$,
${{\dot{\mathbf{ A}}}_r} = \left[ {\frac{{\partial {{\bf{a}}_r}\left( {{\theta _{{T_1}}}} \right)}}{{\partial {\theta _{{T_1}}}}},...,\frac{{\partial {{\bf{a}}_r}\left( {{\theta _{{T_K}}}} \right)}}{{\partial {\theta _{{T_K}}}}},} \right]$, 
${{\dot{\mathbf{ A}}}_t} = \left[ {\frac{{\partial {{\bf{a}}_t}\left( {{\theta _{{T_1}}}} \right)}}{{\partial {\theta _{{T_1}}}}},...,\frac{{\partial {{\bf{a}}_t}\left( {{\theta _{{T_K}}}} \right)}}{{\partial {\theta _{{T_K}}}}}} \right]$, 
and
${\mathbf{R}} = {\mathbf w}{{\mathbf w}^H}$. 	
\begin{proof}
	Please refer to Appendix \ref{sec:appendicesA}.
\end{proof}

The CRB matrix is obtained as
\begin{align}
	{{\bm \Phi }}\left( {{\bm \zeta }} \right)   = \left( {{\mathbf F}\left( {{\bm \zeta }} \right) }\right) ^{-1}.
	\label{crb}
\end{align}

\subsection{Proactive Eavesdropping Model}

It is assumed that the link between $E$ and $D$ is line-of-sight{\footnote
	{	
		This work assumes the links are line-of-sight and the CSI is perfectly known. It is easy to extend the results to scenarios with small-scale fading and/or with imperfect CSI. 
		For the scenarios where there is a non-line-of-sight link between the transmitter and receiver, the channel gains can be approximated by their expectation by utilizing Jensen's inequality, like Refs. \cite{DanQ2025TCCN}, \cite{HuaM2020TCOM}, and \cite{LeiH2024TVT}. 
		For the bounded CSI error model, the distance can be approximated by its bound, which can be obtained based on the triangle inequality \cite{LeiH2024TVT}. 
		For the Gaussian CSI error model, the distances can be approximated by its expectation.
	}
}and the channel coefficient is expressed as \cite{JiaH2023TVT}
\begin{align}
	{{\mathbf h}_{ED}} = \sqrt {\frac{{{\beta _0}}}{{d_{ED}^\alpha }}} {{\mathbf a}_t}\left( {{\theta _D}} \right),
	\label{hed}
\end{align}
where 
$\beta _0$ represents the channel gain at the reference distance,  
$d_{ED}$ represents the distance between $E$ and $D$, 
$\alpha$ denotes the path loss exponent, 
and 
{${\theta _D}$ denotes the angle of $D$}.
Similarly, the channel coefficient between $S$ and $D$ is expressed as
\begin{align}
	{h_{SD}} = \sqrt {\frac{{{\beta _0}}}{{d_{SD}^\alpha }}},
	\label{hsd}
\end{align}
where $d_{SD}$ represents the distance between $S$ and $D$.
The received signal at $D$ is expressed as
\begin{align}
	{{ y}_D} = {h_{SD}}{x_S} + {{\mathbf h}_{ED}^H}{\mathbf x}_E + {n_{D}},
	\label{yd1}
\end{align}
where
$x_S$ denotes the signal sent by $S$,
and
${n_D} \sim N\left( {0,\sigma _D^2} \right)$ is the AWGN.
The SINR of received signal at $D$ is expressed as
\begin{align}
	{\gamma _D} = \frac{{{\beta _0}d_{SD}^{-\alpha }{P_s}}}{{{\beta _0}d_{ED}^{-\alpha }\mathcal{I}\left( {{\theta _D}} \right) + \sigma _D^2}},
	\label{snrD}
\end{align}
where
$P_S$ is the transmit power of $S$
and
$\mathcal{I}\left( {{\theta _D}} \right) = {\mathbf{a}}_t^H\left( {{\theta _D}} \right){\mathbf{R}}{{\mathbf{a}}_t}\left( {{\theta _D}} \right)$ signifies the beampattern gains at $D$ from $E$.

The communication signals received at $E$ is expressed as
\begin{align}
	{\mathbf{y}}_E = {{\mathbf{h}}_{SE}}{x_S} + {\mathbf{ n}}_E,
	\label{yE}
\end{align}
where ${\mathbf{ n}}_E \sim N\left( {0,\sigma _E^2{\mathbf{I}}} \right)$,  ${{\mathbf h}_{SE} \in \mathbb{C}^{N_r\times 1}}$ is the $S$-$E$ channel.
It is assumed that $E$ adopts the selection combining scheme, the SNR of $E$ is expressed as
\begin{align}
	{\gamma _E} = \frac{{{P_S}{{\left| {{h_{SE}}} \right|}^2}}}{{\sigma _E^2}},
	\label{snrE}
\end{align}
where $h_{SE}=\sqrt {\frac{{{\beta _0}}}{{d_{SE}^\alpha }}}$ and $d_{SE}$ represents the distance between $S$ and $E$.
\section{Problem Formulation}
\label{sec:Problem1}

In this section, we design the covariance matrix for PE and TS, respectively, to lay the foundation for optimizing the performance of PE and TS at the same time.

\subsection{Proactive Eavesdropping Only}
\label{Subproblem01}

It is assumed that there is an adaptive information transmission link between $S$ and $D$,
$E$ can decode the received information without error, 	and the success of PE needs to meet ${\gamma _D} \le {\gamma _E}$ \cite{XuD2022SJ, FeiziF2020TCOM}.
In this subsection, the transmitting beam is designed only to meet the eavesdropping demand. 
{To maximize the ER on the premise of successful PE, the following optimization problem is formulated}
\begin{subequations}
	\begin{align}
		\mathcal{P}_{1}:\;  &\mathop {\max }\limits_{\mathbf{R}} {{{\log _2}\left( {1 + {\gamma _D}} \right)}} \nonumber 		\\
		{\mathrm{s.t.}}	
		&	{\mathrm{tr}}({\mathbf{R}}) \le {P_0},    	 	\label{P1a}\\
		&{\mathbf{R}} \succeq 0 ,  								\label{P1b}\\
		&{\mathrm {rank}}\left( {\mathbf{R}} \right) = 1,    		\label{P1c}\\
		&{\gamma _D} \le {\gamma _E},   	\label{P1d}
	\end{align}
\end{subequations}
where
(\ref{P1a}) is the constraint on total power with the maximum allowable power budget $P_0$ of $E$,
(\ref{P1b}) and (\ref{P1c}) are equivalent to ${\mathbf{R}} = {\mathbf w}{{\mathbf w}^H}$  \cite{HeR2025TWC},
(\ref{P1b}) indicates that the covariance matrix $\mathbf{R}$ is a positive semi-definite matrix,
(\ref{P1c}) indicates that the rank of $\mathbf{R}$ is one,
and 
(\ref{P1d}) is the premise of successful eavesdropping. 
Based on (\ref{snrD}) and (\ref{snrE}), (\ref{P1d}) is rewritten as\footnote{
		It is unnecessary for $E$ to send AN to $D$ when the channel quality of $S-E$ is better than that of $S-D$, i.e. $\frac{{h_{SD}^2}}{{\sigma _D^2}} \le \frac{{h_{SE}^2}}{{\sigma _E^2}}$ \cite{JiangX2017SPL}.	
	The ER is maximized only if the interference in the $D$'s direction is zero, in particular, $\mathcal{I}\left( {{\theta _D}} \right)=0 $. Here we mainly consider the nomal case where $\frac{{h_{SD}^2}}{{\sigma _D^2}} > \frac{{h_{SE}^2}}{{\sigma _E^2}}$. }
\begin{align}
	\frac{{d_{ED}^\alpha }}{{{\beta _0}}}\left( {\frac{{h_{SD}^2\sigma _E^2}}{{h_{SE}^2}}-\sigma _D^2} \right) \le \mathcal{I}\left( {{\theta _D}} \right).
	\label{interferenceCons}
\end{align}

Moreover, it is easily found that maximizing ER equals minimizing $\mathcal{I}\left( {{\theta _D}} \right) $, then, $\mathcal{P}_{1}$ is equivalent to
\begin{subequations}
	\begin{align}
		\mathcal{P}_{1.1}:\;  &\mathop {\min }\limits_{\mathbf{R}}  \mathcal{I}\left( {{\theta _D}} \right) \nonumber \\
		{\mathrm{s.t.}}	\,\,\, 
		&(\rm{\ref{P1a}})-(\rm{\ref{P1c}}),(\rm{\ref{interferenceCons}}). \nonumber
	\end{align}
\end{subequations}
$\mathcal{P}_{1.1}$ is challenging to solve because  (\ref{P1c}) is non-convex.
Fortunately, with the proposed method in \cite{SuN2021TWC}, $\mathcal{P}_{1.1}$ can be solved directly by the convex optimization toolbox by ignoring (\ref{P1c}).

\subsection{Target Sensing Only}
\label{Subproblem02}

CRB is the lower bound of the unbiased estimation variance, and therefore minimizing CRB is a strategy for improving the accuracy of the estimation\footnote{{
	There are three different types of scalar CRB metrics based on the trace of CRB, maximum eigenvalue of CRB, and DCRB \cite{HuaH2024TWC, LiJ2008TSP}.
	In this work, the DCRB is chosen as the performance metric of TS. 
	The proposed schemes in this work also fit those scenarios wherein the trace and the maximum eigenvalue of the CRB are utilized as the matrices.} 
}.
In this subsection, the transmitting beam is designed only to minimize the DCRB.

\subsubsection {Perfect Target Direction}
When the given target angle is perfect, the formulated problem is expressed as
\begin{subequations}
	\begin{align}
		\mathcal{P}_{2.1}:\;  &\mathop {\min }\limits_{\mathbf{R}} {\left|{{\bm \Phi }}\left( {{\bm \zeta }} \right)\right|}  	\nonumber \\
		{\mathrm{s.t.}} \,\,\, 	
		&(\rm{\ref{P1a}})-(\rm{\ref{P1c}}), \nonumber \\
		& \mathcal{I}\left( {{\theta _{T_k}}} \right)-\mathcal{I}\left( {{\theta _n}} \right) \ge {\gamma _s},\forall {\theta _n} \in {\Omega},  \forall k,		\label{P2.1d}
	\end{align}
\end{subequations}
where
$\theta _{T_k}$ denotes the direction of $T_k$,
$\theta _n$ prescribes the sidelobe region,
the left-hand side (LHS) of (\ref{P2.1d}) is the
difference between the gain of the target direction and the gain of other directions,
${\Omega}$ represents the angles in the detection range except the target angle $\theta _{T_k}$,
and
$\gamma _s$ is the requirement between the gain of the target direction and the gain of other directions.
The purpose of (\ref{P2.1d}) is to ensure that the gain in the target direction is superior to that in the other directions.
$\mathcal{P}_{2.1}$ is a standard SDR problem and can be solved by CVX toolbox after ignoring the rank-one constraint.

\subsubsection{Imperfect Target Direction}

In this subsection, a more practical scenario is considered wherein the estimated target angle ${\theta _{T_k}}$ is imperfect and there is an uncertain angle $\Delta \theta$.
Specifically, the angle of the $k$-th target varies in
${\Omega_1}=\left[ {{\theta_{T_k}}-\Delta \theta, {\theta_{T_k}} + \Delta \theta } \right]$.
Then, the following optimization problem is formulated
\begin{subequations}
	\begin{align}
		\mathcal{P}_{2.2}:\;   &\mathop {\min }\limits_{\mathbf{R}} {\left|{{\bm \Phi }}\left( {{\bm \zeta }} \right)\right|} 			\nonumber \\
		{\mathrm{s.t.}}\,\,\,
		&(\rm{\ref{P1a}})-(\rm{\ref{P1c}}), \nonumber \\
		& \mathcal{I}\left( {{\theta _{T_k}}} \right)-\mathcal{I}\left( {{\theta _n}} \right) \ge {\gamma _s},\forall {\theta _n} \in {\Omega_2},  						\label{P2.2d}\\
		& \mathcal{I}\left( {{\theta _m}} \right) \le \left( {1 + \varphi } \right) \mathcal{I}\left( {{\theta _{T_k}}} \right), \forall {\theta _m} \in {\Omega _1}, 		\label{P2.2e}\\
		& \mathcal{I}\left( {{\theta _m}} \right) \ge \left( {1-\varphi } \right) \mathcal{I}\left( {{\theta _{T_k}}} \right),  \forall {\theta _m} \in {\Omega _1}, 		\label{P2.2f}
	\end{align}
\end{subequations}
where 
$\Omega_2$ represents other angles except the mainlobe and 
$0 \leq \varphi \leq 1$ is a given scalar associated with the wide main beam fluctuation \cite{SuN2024TWC}, 
(\ref{P2.2d}) represents the gain gap constraint between the target direction and side lobes,
(\ref{P2.2e}) and (\ref{P2.2f}) are both stable mainlobe gain intervals to prevent a large gain difference which affects the detection performance.
$\mathcal{P}_{2.2}$ can be solved by CVX after ignoring (\ref{P1c}).

\section{Jointly Optimization of Proactive Eavesdropping and TS}
\label{sec:Problem3}	
	
Notice that in the considered system, the transmitted sensing signal is utilized as AN to interfere with the illegitimate receivers, like the works on PLS of the ISAC systems \cite{RenZ2024TWC}, \cite{SuN2021TWC}, \cite{SuN2022TWC}, \cite{WeiW2024TWC}, \cite{DongF2023TGCN},  \cite{RenZ2023TCOM}, \cite{SuN2024TWC}.
In these outstanding works, the more interference the illegitimate receivers, the better the secrecy performance. 
However, in our work, the sensing signals need to be designed carefully. 
Too weak interference to the illegitimate receivers will result in the failure of PE, while excessive interference to the illegitimate receivers will result in lower eavesdropping rates. 
Based on the problems formulated in Section III, the PE-CRB region characterizes the achievable region for the target parameter estimation and PE is defined as
\begin{align}
	{\mathbb{PS}} &= \bigcup\limits_{{\mathbf{R}} \succeq 0} \left\{ { {{{\log }_2}\left( {1 + {\gamma _D}} \right)},  {\left| {{\bm \Phi} \left( {\bm \zeta}  \right)} \right|} } {\left| {{\gamma _D} < {\gamma _E},\left| {{\bm \Phi} \left( {\bm \zeta}  \right)} \right| \le {\alpha _\Phi }} \right.} \right\},
	\label{region}
\end{align}
where $\alpha_{\Phi}$ denotes the target threshold for determinant of the CRB.

\subsection{Proactive Eavesdropping-centric Design}

Firstly, the transmit covariance matrix is optimized to minimize the beampattern gains at $D$ while ensuring a target threshold for the estimation CRB for multi-target constraint into account (called as PE-centric design). 
Based on $\mathcal{P}_{1.1}$, the optimization problem is formulated as 	
\begin{subequations}
	\begin{align}
		\mathcal{P}_{3}:\;  &\mathop {\min }\limits_{\mathbf{R}}  \mathcal{I}\left( {{\theta _D}} \right) \nonumber \\
		{\mathrm{s.t.}}	\,\,\, 
		&(\rm{\ref{P1a}})-(\rm{\ref{P1c}}),(\rm{\ref{interferenceCons}}), \nonumber \\
		&{\left|{{\bm \Phi }}\left( {{\bm \zeta }} \right)\right|} \leq \alpha_{\Phi},	\label{P3CRB}
	\end{align}
\end{subequations}	
where 
(\ref{P3CRB}) is the TS constraint and $\alpha_{\Phi}$ denotes the target threshold for determinant of the CRB. 
$\mathcal{P}_{3}$ is a convex problem after dropping constraint (\ref{P1c}), which can be solved by CVX toolbox.

\subsection{Target Sensing-centric Design}

Then, we consider optimizing the transmit covariance matrix to minimize the estimation CRB while ensuring a beampattern gains (interference) at $D$ while taking the maximum transmit power constraint into account (called as TS-centric design). 
Based on $\mathcal{P}_{2.1}$ and $\mathcal{P}_{2.2}$, the optimization problem is formulated as 		
\begin{subequations}
	\begin{align}
		\mathcal{P}_{4.1}:\;  &\mathop {\min }\limits_{\mathbf{R}} {\left|{{\bm \Phi }}\left( {{\bm \zeta }} \right)\right|}  	\nonumber \\
		{\mathrm{s.t.}} \,\,\, 	
		&(\rm{\ref{P1a}})-(\rm{\ref{P1c}}),(\rm{\ref{interferenceCons}}), (\rm{\ref{P2.1d}}),\nonumber \\
		&\mathcal{I}\left( {{\theta _D}}\right) \le \alpha_{I}, \label{P4.1f}
	\end{align}
\end{subequations}
and 
\begin{subequations}
	\begin{align}
		\mathcal{P}_{4.2}:\;  &\mathop {\min }\limits_{\mathbf{R}} {\left|{{\bm \Phi }}\left( {{\bm \zeta }} \right)\right|}  	\nonumber \\
		{\mathrm{s.t.}} \,\,\, 	
		&(\rm{\ref{P1a}})-(\rm{\ref{P1c}}),(\rm{\ref{interferenceCons}}), 
		(\rm{\ref{P2.2d}})-(\rm{\ref{P2.2f}}),(\rm{\ref{P4.1f}}). \nonumber 
	\end{align}
\end{subequations}
where 
$\alpha_{I}$ denotes the interference threshold at $D$. 
It must be noted that (\ref{P4.1f}) denotes the requirement for the ER.
Obviously, $\alpha_{I}$ has an essential effect on the optimized results.
$\mathcal{P}_{4.1}$ and $\mathcal{P}_{4.2}$ can be transformed into a convex problem by means of SDR and are solved directly using the CVX toolbox when (\ref{P1c}) is omitted.

\color{black}
\subsection{Normalized Weighted Optimization of PE and Target Sensing.}

{In this subsection, a normalized weighted optimization problem that simultaneously considers both PE and TS performance is proposed.}
Design of the weighted optimization between the CRB and ER presents the challenge that both performance metrics have different units and potentially different magnitudes.
Inspired by \cite{SuN2024TWC}, both $\mathcal{I}\left( {{\theta _D}} \right) $ and ${\left|{{\bm \Phi }}\left( {{\bm \zeta }} \right)\right|}$ are normalized with their respective bound obtained in $\mathcal{P}_{1.1}$ and $\mathcal{P}_{2.1}$, respectively.

\subsubsection {Joint Optimization Proactive Eavesdropping and Target Sensing}

{To simultaneously consider PE and TS, }while taking the uncertain angle of $T_k$ into account, the weighted optimization problem is formulated as\footnote{{
	It should be noted that (\ref{P3CRB}) and (\ref{P4.1f}) correspond to $\rho = 1$ and $\rho = 0$, respectively. In particular, $\mathcal{P}_{5.1}$ degenerates into $\mathcal{P}_{3}$ and $\mathcal{P}_{4.2}$ when when $\rho = 1$ and $\rho = 0$, respectively.
}}
\begin{subequations}
	\begin{align}
		\mathcal{P}_{5.1}:\;  &	\mathop {\min }\limits_{\mathbf{R}} \rho \frac{{\mathcal{I}\left( {{\theta _D}} \right)}}{{ {{\mathrm{\tilde {\mathcal{I}}}}} }} + \left( {1-\rho } \right)\frac{{\left|{{\bm \Phi }}\left( {{\bm \zeta }} \right)\right|} }{{\left| {{{\tilde {\bm \Phi   }}}} \right|}} 		\nonumber \\
		{\mathrm{s.t.}}\,\,\, 
		&(\rm{\ref{P1a}})-(\rm{\ref{P1c}}),(\rm{\ref{interferenceCons}}), \nonumber \\
		&(\rm{\ref{P2.2d}})-(\rm{\ref{P2.2f}}), (\rm{\ref{P3CRB}}), (\rm{\ref{P4.1f}}). \nonumber  										
	\end{align}
\end{subequations}
where
$\mathcal{\tilde I} $ and ${{{\tilde {\bm \Phi}}}}$ are the results of $\mathcal{P}_{1.1}$ and $\mathcal{P}_{2.1}$ $\left( \mathcal{P}_{2.2} \right)$, respectively,
and
$0 \leq \rho  \leq 1$ signifies the weights factor between PE and TS. $\mathcal{P}_{5.1}$ is an SDR problem by neglecting (\ref{P1c}) and can be solved with the CVX toolbox. 	
	
\subsubsection {The interference power is zero}

It is unnecessary for $E$ to send AN to $D$ when the channel quality of $S-E$ is better than that of $S-D$, i.e. $\frac{{h_{SD}^2}}{{\sigma _D^2}} \le \frac{{h_{SE}^2}}{{\sigma _E^2}}$ in (\ref{interferenceCons}) \cite{JiangX2017SPL}.
In these scenarios, the ER is equal to the receiving rate of $D$ without interference and $\mathcal{P}_{3}$ is transformed into a TS sub-problem, which is expressed as
\begin{subequations}
	\begin{align}
		\mathcal{P}_{5.2}:\;  & \mathop {\min }\limits_{\mathbf{R}}  {\left|{{\bm \Phi }}\left( {{\bm \zeta }} \right)\right|} 	\nonumber \\
		{\mathrm{s.t.}} \,\,\, 	
		&(\rm{\ref{P1a}})-(\rm{\ref{P1c}}),(\rm{\ref{interferenceCons}}), (\rm{\ref{P2.2d}})-(\rm{\ref{P2.2f}}), \nonumber \\
		& {{ \mathcal{I}\left( {{\theta _D}} \right) \leq \epsilon }}, 				\label{P3.1d}
	\end{align}
\end{subequations}
{where $\epsilon$ is a constant.
	By selecting an appropriate $\epsilon$, the eavesdropping performance can approximate the ideal scenario with zero interference, thereby striking a balance between engineering practicality and theoretical performance.}

$\mathcal{P}_{5.2}$ is a standard convex problem when (\ref{P1c}) is omitted, which can be solved directly by CVX. 	
The solving process of $\mathcal{P}_{5.1}$ ($\mathcal{P}_{5.2}$) ignoring rank-one constraint is summarized as \textbf{Algorithm 1}.

\begin{algorithm}[t]
	{
	\caption{Joint Optimization of TS and PE without Rank-one Constraint}
	\KwIn{ 
		Initialize $P_0$, $\Delta \theta$, $\varphi$, $\gamma_s$, $\gamma_D$, and $\gamma_E$.\\
	}
	\If{$\gamma_D > \gamma_E$	}
	{
		Solve $\mathcal{P}_{1.1}$ for given parameters and obtain the solution ${ {{\mathrm{\tilde {\mathcal{I}}}}} }\left( {{\theta _D}} \right)$.\\
		Solve $\mathcal{P}_{2.1}$ for given parameters and obtain the solution ${\left|{{\bm {\tilde \Phi} }}\right|}$. \\
		Solve $\mathcal{P}_{5.1}$  for the given parameters and obtain the solution.
	}
	\Else{
		Solve $\mathcal{P}_{5.2}$ for the given parameters and obtain the solution.
	}
	\KwOut{The covariance matrix $\mathbf{R}$ of $E$.}
}
\end{algorithm}

\subsection{Rank-one Constraint}

The rank-one constraint is non-convex, which means the relevant toolbox can not directly solve the optimization problem. The above issues are solved by omitting the rank-one constraint and the rank of the obtained solution may be larger than one. 	
Based on Refs. \cite{SuN2021TWC} and \cite{SuN2024TWC}, the feasible suboptimal solution of the original problem is obtained by utilizing specific methods based on the SDR technology, and rank-one constraint is solved by Gaussian randomization, singular value decomposition, etc. It must be noted that these methods may result in a certain degree of distortion, which is tolerable in many cases. However, these methods cannot be utilized to solve the previously formulated problems since the recovered result may not satisfy ${\gamma _D} \le {\gamma _E}$, which results in eavesdropping failure.
{
	In this work, an iterative method, SROCR, is utilized to solve this problem \cite{CaoP2017EUSIPCO, LiuZ2023TVT}. Specifically, the rank-one constraint is relaxed by restricting the ratio of the maximum eigenvalue and trace of the matrix.
}	
In particular, for the scenario with ${\mathrm{rank}}\left( {\mathbf{\tilde R}} \right) > 1$, the eigenvalue decomposition of the matrix is firstly carried out and the eigenvalue matrix ${\mathbf{\tilde \Sigma}} = {\mathrm {diag}} \left({{\lambda _1}, ..., {\lambda _{{N_t}}}} \right)$ is obtained. 
According to the relationship between the matrix eigenvalue and the matrix trace, the maximum eigenvalue of  ${\mathbf{\tilde R}}$ must be less than the trace. The eigenvector of maximum eigenvalue of ${\mathbf{\tilde R}}$ is expressed as ${\mathbf u}$ and there must be ${{\mathbf u}^H}{\mathbf{\mathbf{\tilde R}}}{\mathbf u} < {\mathrm {tr}}\left( {{\mathbf{\mathbf{\tilde R}}}} \right)$. By using the iterative method, the one constraint is transformed into
\begin{align}
	{{\mathbf u}^{\left( {j} \right)}}^H{\mathbf{R}}{{\mathbf u}^{\left( {j} \right)}} \ge {w^{\left( {j } \right)}}{\mathrm{tr}}\left( {\mathbf{R}} \right),
\end{align}
where ${\left(  \cdot  \right)^{\left( j \right)}}$ denotes $j$-th iteration value and $w \in [0,1]$ is the relaxation parameter.
	
Then $\mathcal{P}_{5.1}$ and $\mathcal{P}_{5.2}$ are reformulated as
\begin{subequations}
	\begin{align}
		\mathcal{P}_{5.{\mathrm {1b}}}:\;  &\mathop {\min }\limits_{\mathbf{R}} \rho \frac{{\mathcal{I}\left( {{\theta _D}} \right)}}{{ {{\mathrm{\tilde {\mathcal{I}}}}} }} + \left( {1-\rho } \right)\frac{{\left|{{\bm \Phi }}\left( {{\bm \zeta }} \right)\right|} }{{\left| {{{\tilde {\bm \Phi   }}}} \right|}} \nonumber\\
		{\mathrm{s.t.}}\,\,\,
		&(\rm{\ref{P1a}}), (\rm{\ref{P1b}}),(\rm{\ref{interferenceCons}}), (\rm{\ref{P2.2d}})-(\rm{\ref{P2.2f}}), (\rm{\ref{P3CRB}}), (\rm{\ref{P4.1f}}), \nonumber \\
		&{{\mathbf u}^{\left( {j} \right)}}^H{\mathbf{R}}{{\mathbf u}^{\left( {j} \right)}} \ge {w^{\left( {j } \right)}}{\mathrm{tr}}\left( {\mathbf{R}} \right), 	\label{P5.1bnew}
	\end{align}
\end{subequations}
and
\begin{subequations}
	\begin{align}
		\mathcal{P}_{5.{\mathrm {2b}}}:\;  &\mathop {\min }\limits_{\mathbf{R}} {\left|{{\bm \Phi }}\left( {{\bm \zeta }} \right)\right|} \nonumber\\
		{\mathrm{s.t.}}\,\,\,
		&(\rm{\ref{P1a}}),(\rm{\ref{P1b}}),(\rm{\ref{interferenceCons}}), (\rm{\ref{P2.2d}})-(\rm{\ref{P2.2f}}), (\rm{\ref{P5.1bnew}}), \nonumber	 
	\end{align}
\end{subequations}
respectively,
where 
$ {{w^{\left( 0 \right)}}} = \frac{{{{\mathbf u}^H}{\mathbf{\mathbf{\tilde R}}}{\mathbf u}}}{{{\mathrm {tr}}\left({\mathbf{\tilde R}} \right)}} \in \left[ {0,1} \right]$ and ${\mathbf{\tilde R}}$
is obtained by solving $\mathcal{P}_{5.1}$ $(\mathcal{P}_{5.{\mathrm 2}})$. 
The optimal result of $\mathcal{P}_{5.{\mathrm {1b}}}$ $(\mathcal{P}_{5.{\mathrm {2b}}})$ is denoted as $opt^{(j)}$ in the $j$-th iteration,
which are solved by an iterative algorithm summarized as \textbf{Algorithm 2}, where $\vartheta$ denotes the predetermined accuracy. 
After obtain $\mathbf{R}$ by solving $\mathcal{P}_{5.{\mathrm {1b}}}$ $(\mathcal{P}_{5.{\mathrm {2b}}})$, the beamforming vector, ${\mathbf w}$, is obtained through the eigenvalue decomposition.

\begin{algorithm}[t]
	{
	\caption{Iterative Procedure of $\mathcal{P}_{5.1}$ ($\mathcal{P}_{5.{\mathrm {2b}}}$) with Rank-one Constraint}
	\KwIn{ Initialize $P_0$, $\Delta \theta$, $\varphi$, $\gamma_s$, $\gamma_D$,$ \gamma_E$, $\tau$, and $\vartheta$.}
	$j \leftarrow 0$.\\
	Solve $\mathcal{P}_{5.1}$ ($\mathcal{P}_{5.2}$) and obtain ${\mathbf R}$.\\
	${w^{(0)}} \leftarrow \frac{{{{\mathbf u}^H}\mathbf R{\mathbf u}}}{{{\mathrm {tr}}\left( \mathbf R \right)}}$.\\
	${\delta ^{(0)}} \leftarrow \max\left[ {0,1-\frac{{{{\mathbf u}^H}{\mathbf{R}}{{\mathbf u}}}}{{{\mathrm {tr}}\left( {\mathbf{R}} \right)}}} \right]$.\\
	\While{$\| opt^{(j+1)}-opt^{(j)}\| \ge \vartheta$ and $w^{\left( {j} \right) }\le \tau$}
	{
		Solve $\mathcal{P}_{5.{\mathrm {1b}}}$ ($\mathcal{P}_{5.{\mathrm {2b}}}$).\\
		\If{$\mathcal{P}_{5.{\mathrm {1b}}}$ ($\mathcal{P}_{5.{\mathrm {2b}}}$) is solvable}
		{
			${{\mathbf{R}}^{\left( {j + 1} \right)}}$ and ${\delta ^{\left( {j + 1} \right)}} = {\delta ^{\left( {j} \right)}}$.
		}
		\Else{
			${{\mathbf{R}}^{\left( {j + 1} \right)}} = {{\mathbf{R}}^{\left( j \right)}}$ and ${\delta ^{\left( {j + 1} \right)}} = \frac{1}{2}{\delta ^{\left( {j} \right)}}$.
		}
		${w^{\left( {j + 1} \right)}} \leftarrow \min \left( {\frac{{{{\mathbf u}^{\left( {j+1} \right)}}^H{{\mathbf{R}}^{\left( {j + 1} \right)}}{{\mathbf u}^{\left( {j + 1} \right)}}}}{{{\mathrm {tr}}\left( {{{\mathbf{R}}^{\left( {j + 1} \right)}}} \right)}} + {\delta ^{\left( {j + 1} \right)}},1} \right)$.\\
		$j \leftarrow j + 1$.\\
	}

	\KwOut{${{\mathbf{R}}^{\left( {j} \right)}}$}
}
\end{algorithm}

\subsection{Comparison of formulated problems}
In this work, there are eight (two kinds of) formulated problems, which are introduced as follows. 

\begin{itemize}
	\item The first kind of problem is considering PE or TS, respectively.
	\begin{enumerate}
		\item In $\mathcal{P}_{1}$ ($\mathcal{P}_{1.1}$), the transmitting beam is designed only to maximize the ER, considering successful PE, and the TS is not considered.
		
		\item In $\mathcal{P}_{2.1}$ and $\mathcal{P}_{2.2}$, the transmitting beam is designed only to minimize the DCRB and the PE is not considered. $\mathcal{P}_{2.1}$ assumes that the azimuth angle of targets (${\theta _{T_k}}$) are perfect, while $\mathcal{P}_{2.2}$ examines the imperfect ${\theta _{T_k}}$.
		
		\item The results of $\mathcal{P}_{1}$ and $\mathcal{P}_{2.2}$ are utilized in $\mathcal{P}_{5.1}$ and $\mathcal{P}_{5.2}$.
		
	\end{enumerate}
	
	\item The second kind of problem is considering PE and TS simultaneously.
	\begin{enumerate}
		\item In $\mathcal{P}_{3}$, the transmit covariance matrix is optimized to maximize the PE performance (minimize the beampattern gains at $D$) while taking the TS constraint into account, which is called PE-centric (PEC) design.
		
		\item In $\mathcal{P}_{4.1}$ and $\mathcal{P}_{4.2}$, the transmit covariance matrix is optimized to maximize the TS performance (minimize the estimation CRB) while ensuring successful PE and ER constraints, which is called the TS-centric (TSC) design. \textit{$\mathcal{P}_{4.1}$ assumes that the azimuth angle of targets (${\theta _{T_k}}$) is perfect, while $\mathcal{P}_{4.2}$ examines the imperfect ${\theta _{T_k}}$.}
		
		\item In $\mathcal{P}_{5.1}$ and $\mathcal{P}_{5.2}$, the normalized weighted summation between ${\left|{{\bm \Phi }}\left( {{\bm \zeta }} \right)\right|}$ and $\mathcal{I}\left( {{\theta _D}} \right)$ is minimized considering successfully PE, DCRB requirements, interference requirement at $D$, which is called as joint optimization of PE and TS (JPT) design. 
		\textit{Different from $\mathcal{P}_{5.1}$, $\mathcal{P}_{5.2}$ considers a special scenario wherein the channel quality of $S-E$ is better than that of $S-D$, i.e. $\frac{{h_{SD}^2}}{{\sigma _D^2}} \le \frac{{h_{SE}^2}}{{\sigma _E^2}}$. In this case, $E$ doesn't have to send AN to $D$, and all the power can be utilized to TS.}		
	\end{enumerate}	
\end{itemize}	
Table \ref{table3} summarizes the difference for all the formulated problems.
\begin{table}
	\centering
	\caption{Comparison of formulated problems.}
	\label{table3}
	\begin{threeparttable}
		\resizebox{0.46\textwidth}{!}{
			\begin{tabular}{c|c|c|c|c}
				\Xhline{1.2pt}
				{{\makecell[c]{Problem}}} & \textbf{PE} & {\makecell[c]{TS}} &{\makecell[c]{Perfect/imperfect \\Target Direction}}  & {$\frac{{h_{SD}^2}}{{\sigma _D^2}} \le \frac{{h_{SE}^2}}{{\sigma _E^2}}$}\\
				\hline				
				$\mathcal{P}_1$ ($\mathcal{P}_{1.1}$) &\checkmark&  &   &  \\
				\hline
				$\mathcal{P}_{2.1}$ &   & \checkmark &  Perfect &   \\
				\hline
				$\mathcal{P}_{2.2}$ &   & \checkmark &  Imperfect &  \\
				\hline
				$\mathcal{P}_{3}$ &   \checkmark & Constraint &   &  \\
				\hline
				$\mathcal{P}_{4.1}$ & Constraint   & \checkmark &  Perfect &  \\	
				\hline
				$\mathcal{P}_{4.2}$ & Constraint   & \checkmark &  Imperfect &  \\
				\hline
				$\mathcal{P}_{5.1}$ ($\mathcal{P}_{5.{\mathrm {1b}}}$) & \checkmark  & \checkmark &  Imperfect &   \\
				\hline
				$\mathcal{P}_{5.2}$ ($\mathcal{P}_{5.{\mathrm {2b}}}$) & \checkmark  & \checkmark &  Imperfect & \checkmark \\ 					
				\Xhline{1.2pt}
		\end{tabular}}
	\end{threeparttable}
\end{table}

\subsection{Convergent Discussion and Complexity Analysis}

While optimization problems involving multi-variable coupling often necessitate iterative approaches such as BCD or SCA. The convergence analysis serves as key theoretical validation. This work formulates problems containing only a single variable, \(\mathbf{R}\). By applying SDR to relax the rank-one constraint, the problem is transformed into a convex optimization task that can be solved directly. Although this relaxation reduces computational complexity, it may introduce a performance gap between the relaxed solution and the true optimum of the original non-convex problem. 		
To mitigate this gap and enhance solution tightness, we utilize an iterative optimization method based on rank-one relaxation. Specifically, we define the ratio of the maximum eigenvalue to the matrix trace as a relaxation performance metric and adopt a binary search strategy to refine the solution while ensuring constraint feasibility iteratively. Crucially, this approach guarantees strict improvement in the relaxation metric with each iteration. By setting a convergence threshold (e.g., 0.999), the final solution's performance index is driven arbitrarily close to 1, achieving a high-precision approximation of the original problem's optimum. This method not only circumvents the computational challenges of directly solving non-convex problems but also significantly improves accuracy through iterative refinement, offering a practical and theoretically grounded solution pathway for rank-one-constrained non-convex optimization.

The complexity analysis is as follows. 
$\mathcal{P}_{1.1}$ is composed of 1 linear matrix inequality (LMI) constraint of size $N_t$ and 2 LMI constraints of size 1, the computational complexity is $O\left( {\sqrt {{N_t} + 2} \left[ {N_t^2\left( {N_t^3 + 2} \right) + N_t^4\left( {N_t^2 + 2} \right) + N_t^6} \right]} \right)$. $\mathcal{P}_{2.2}$ is composed by 1 LMI constraints of size $N_t$, ${2K{\varpi _1} + K{\varpi _2} + 1}$ LMI constraints of size 1,  the computational complexity is $O\left( \sqrt {{N_t} + \left( {2K{\varpi _1} + K{\varpi _2} + 1} \right)} \left[ N_t^2\left( {N_t^3 + 2K{\varpi _1} + K{\varpi _2} + 1} \right) +\right.\right.$\\ $\left.\left. N_t^4  \left( {N_t^2 + 2K{\varpi _1} + K{\varpi _2} + 1} \right) + N_t^6 \right] \right)$, where ${\varpi _1}$ and ${\varpi _2}$ are cardinality of ${\Omega _1}$ and ${\Omega _2}$, respectively.
$\mathcal{P}_{5.1}$ have 1 LMI constraints of size $N_t$, ${2K{\varpi _1} + K{\varpi _2} + 4}$ LMI constraint of size 1, 
$\mathcal{P}_{5.2}$ have 1 LMI constraints of size $N_t$, ${2K{\varpi _1} + K{\varpi _2} + 2}$ LMI constraint of size 1, 
$\mathcal{P}_{\mathrm {5.{\mathrm {1b}}}}$  have 1 LMI constraints of size $N_t$, ${2K{\varpi _1} + K{\varpi _2} + 5}$ LMI constraint of size 1, 
$\mathcal{P}_{5.{\mathrm {2b}}}$ have 1 LMI constraints of size $N_t$, ${2K{\varpi _1} + K{\varpi _2} + 3}$ LMI constraint of size 1.
With the given convergence accuracy $\vartheta$, the computational complexity of algorithm 1 is expressed as $O\left( {\sqrt {{N_t} + 2K{\varpi _1} + K{\varpi _2}} N_t^6 \log \left( {\frac{1}{\vartheta }} \right)}  \right)$ \cite{SuN2022TWC, SuN2024TWC}, the computational complexity of algorithm 2 is $O\left(  {{I_{{\mathrm{iter}}}}} \left( \sqrt {{N_t} + 2K{\varpi _1} + K{\varpi _2}} N_t^6 \right)\log \left( {\frac{1}{\vartheta }} \right) \right)$, while $I_{\mathrm{iter}}$ represents the number of iterations.

\begin{figure*}[t]
	\centering
	\subfigure[Beampattern gains of JPT with varying $P_0$.]{
		\label{fig02a}
		\includegraphics[width = 0.3  \textwidth]{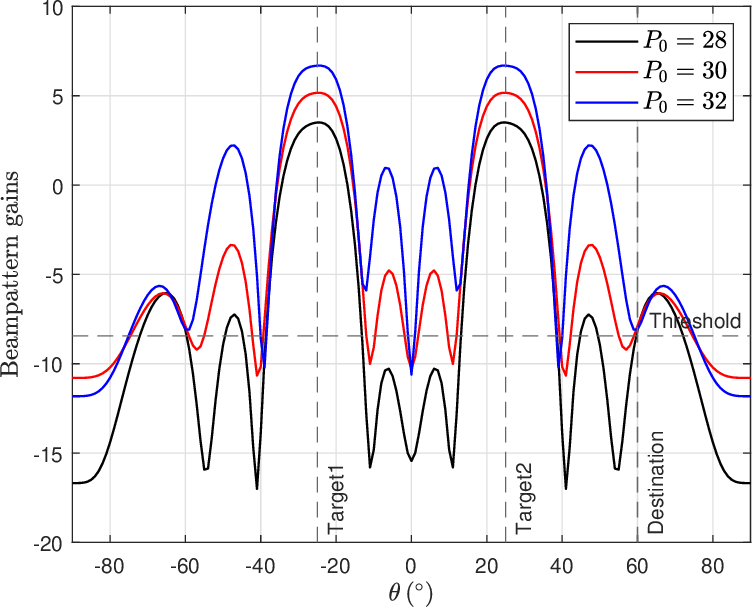}}
	\subfigure[CRB versus varying $P_0$.]{
		\label{fig02b}
		\includegraphics[width = 0.3  \textwidth]{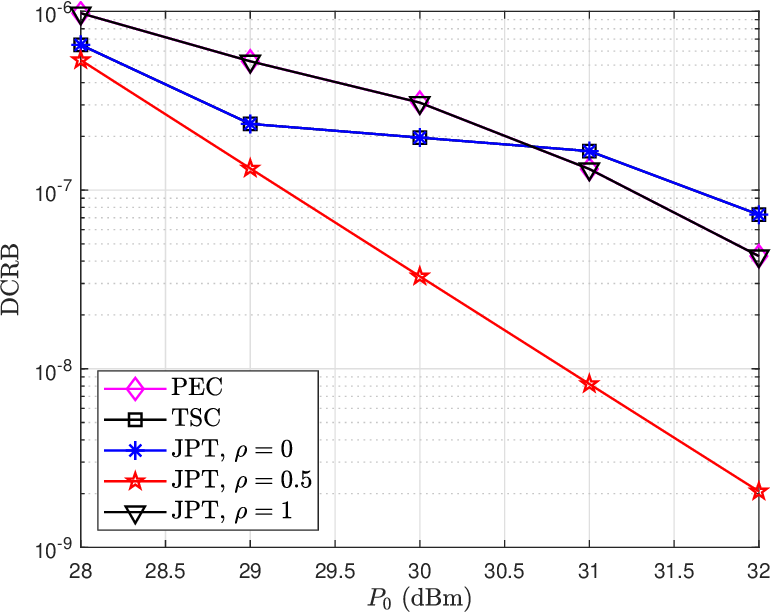}}
	\subfigure[ER versus varying $P_0$.]{
		\label{fig02c}
		\includegraphics[width = 0.3  \textwidth]{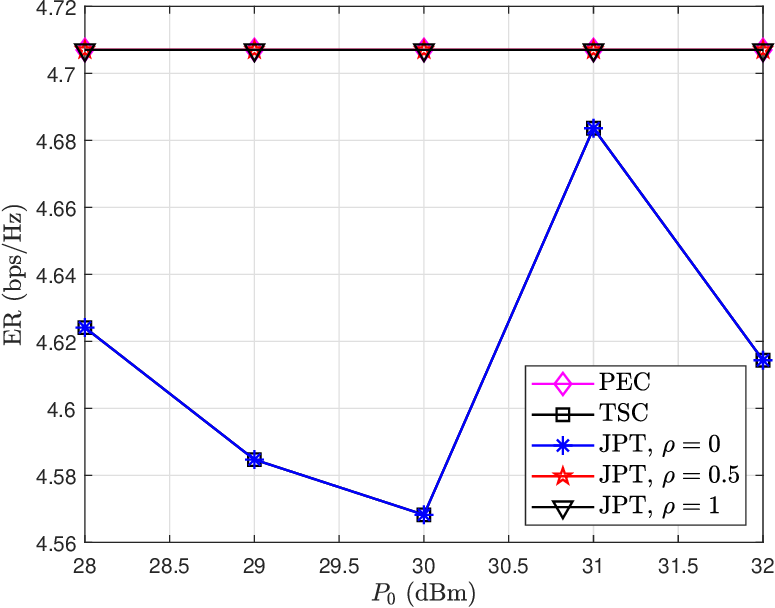}}
	\caption{{The results versus varying $P_0$ with $N_t = 12$ and $\Delta \theta = 3$.}}
	\label{fig02}
\end{figure*}
\section{Numerical Results and Discussion}
\label{sec:Simulation}

In this section, numerical results are presented to prove the convergence and effectiveness of our algorithm. 
The positive direction of the X axis is used as the reference, and the azimuth angles of $T_k$ $\left( {K = 2} \right)$ and $D$ relative to $E$ are expressed as ${\theta _{T_k}}$ and ${\theta _D}$, respectively.
The direction of the radar detection range is $\left[ {-{{90}^ \circ },{{90}^ \circ }} \right]$ , $\sigma^2$ is normalized to be unity \cite{HuaH2024TWC}, \cite{LiJ2008TSP},
and other parameters are listed in Table \ref{table4}.

\begin{table}
	\centering
	\caption{\textit{Parameters Setting}}
	\begin{tabular}{c|c|c|c}
		\Xhline{1.2pt}
		Parameters    &   Value            &    Parameters    &   Value      \\
		\hline
		${d}_{SD}$ &  500 m & ${d}_{SE} $ &1000 m\\
		\hline
		${d}_{ED}$ & 1000 m& $\theta _{T}$	   & $-25^ \circ/25^ \circ$ \\
		\hline
		$\theta _D $	&  $60^ \circ$ & $\beta _0$				& $-30$ dB	 			\\
		\hline
		$\lambda$ &  0.06 m  &  $d$ & 0.03m \\
		
		\hline
		$ \sigma_D^2=\sigma_E^2$				& $-80$ dBm 			& $\varphi$                  & 0.05       \\
		\hline
		$\gamma _s$ &0.1 & $\alpha $					& 2.2					\\
		\hline
		$P_S$	& 30 dBm & $\vartheta$				& $10^{-4}$					\\ 	
		\hline
		$\tau $	&  0.999  & $N_r = N_t $ 					& 12	 			\\ 	
		\hline
		$\alpha_{\Phi}$	&  $10^{-5} $ & $\alpha_{I} $ 					& 0.2	 			\\ 	
		\Xhline{1.2pt}
	\end{tabular}
	\label{table4}
\end{table}

\begin{figure*}[t]
	\centering
	\subfigure[Beampattern gains of JPT with $P_0 = 30$ dBm.]{
		\label{fig03a}
		\includegraphics[width = 0.3  \textwidth]{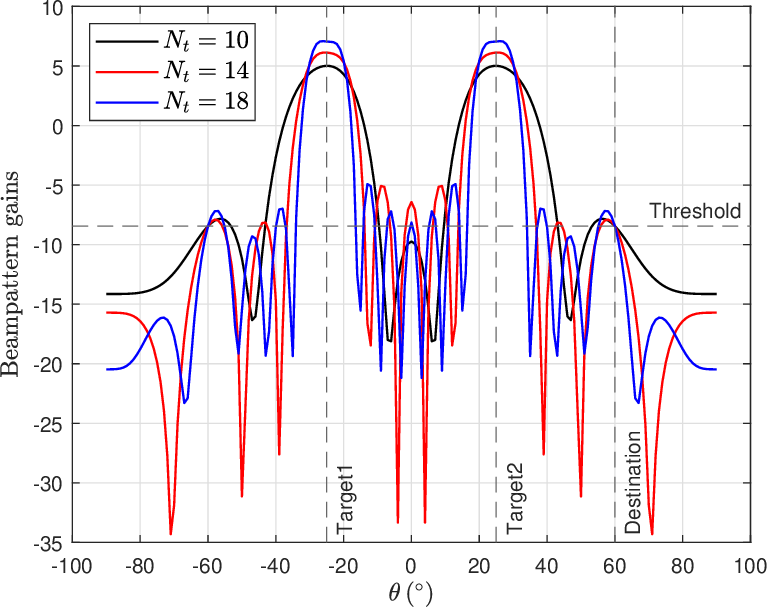}}
	\subfigure[DCRB versus varying $N_t$.]{
		\label{fig03b}
		\includegraphics[width = 0.3  \textwidth]{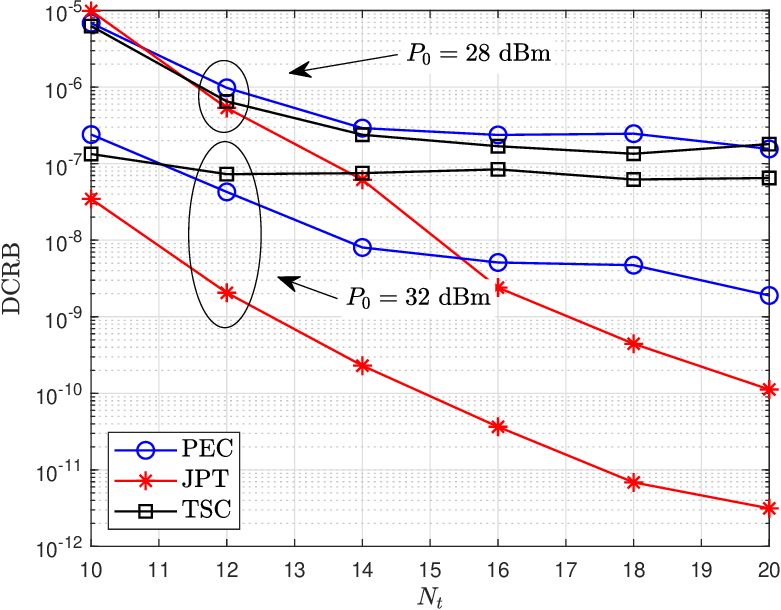}}
	\subfigure[ER for varying $N_t$.]{
		\label{fig03c}
		\includegraphics[width = 0.3  \textwidth]{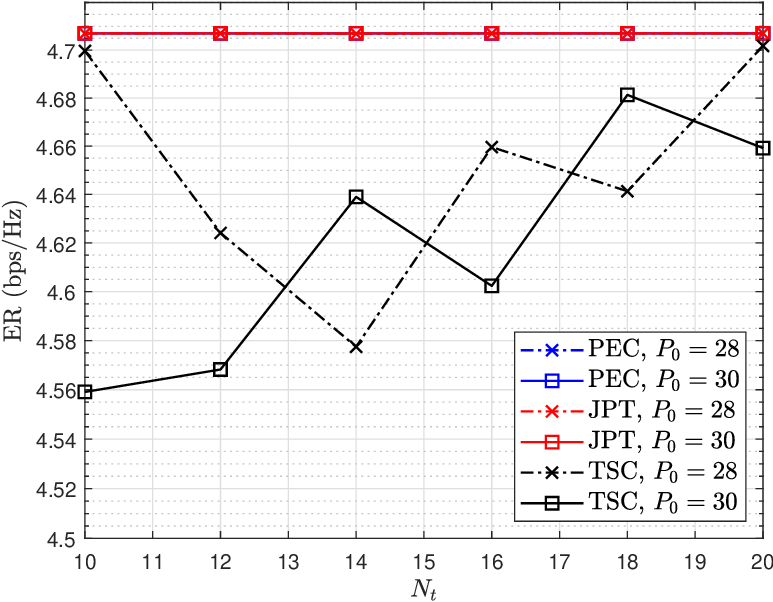}}
	\caption{{The results versus varying $N_t$ with $\rho = 0.5$ and $\Delta \theta = 3$.}}
	\label{fig03}
\end{figure*}

Fig. \ref{fig02a} plots the beampattern gains of the JPT scheme with varying power budgets.
It can be observed that the gains of the mainlobe increases with increasing $P_0$, while the width of the mainlobes remains unchanged.
Fig. \ref{fig02b} demonstrates the relationship between CRB and radar power budget for various schemes.
It can be observed that the DCRB decreases with the increase of $P_0$, which denotes that increasing the power budget can improve the accuracy of sensing. 
When $\rho=0$, JPT is equivalent to the TSC scheme; when $\rho=1$, JPT is equivalent to the PEC scheme.		
The JPT scheme's sensing performance outperforms that of the TSC scheme. 
This is because the beampattern gains in $D$'s direction is limited in the JPT scheme. With the increasing budget, more power is allocated to the mainlobe, enhancing the sensing performance. 
Fig. \ref{fig02c} plots the ER of various schemes versus the varying $P_0$. 
We can find that both the PEC and JPT ($\rho \ne 0$) schemes obtain the maximum ER. 
The ER of the TSC and JPT with $\rho = 0$ are the worst exhibits no significant regular variation with the increase of $P_0$. 
That is because (\ref{interferenceCons}) and (\ref{P4.1f}) correspond to successful PE and the maximum interference at $D$, respectively and no maximizing ER in TSC and JPT with $\rho = 0$.
Obviously, larger $\alpha_{I}$, less ER due to $\alpha_{I}$ denotes the interference threshold at $D$.

Fig. \ref{fig03a} shows the relationship between the beampattern gains of the JPT scheme and $N_t$.
The results show that the mainlobe's beampattern gains increases and its width decreases with increasing $N_t$. 
Fig. \ref{fig03b} demonstrates the sensing performance of all the schemes is improved with increasing $N_t$. 
Specifically, when the number of antennas is small, the TSC scheme's sensing performance is better than that of the JPT scheme. 
As the number of antennas increases, the JPT scheme's sensing performance increases faster than that of the TSC scheme.
Moreover, the PEC scheme's sensing performance tends to be constant, which outperforms that of the TSC scheme since the beampattern gains in $D$'s direction is optimized so that more power is allocated to the mainlobe in the PEC scheme. 
{{
		Fig. \ref{fig03c} plots the ER for all the schemes versus varying $N_t$. 
		We can see that both the PEC and JPT schemes achieve the maximum ER (the achievable rate of $S$-$D$), which is independent of the number of antennas on $E$. 
		Because only successful PE is guaranteed in the TSC scheme, which is characterized by whether the minimum beam pattern gains in the direction of $D$ are satisfied. Therefore, the ER in the TSC scheme has a certain degree of randomness.
		 }}

\begin{figure*}[t]
	\centering
	\subfigure[Beampattern gains for different schemes with $P_0 = 30$ dBm.]{
		\label{fig04a}
		\includegraphics[width = 0.3  \textwidth]{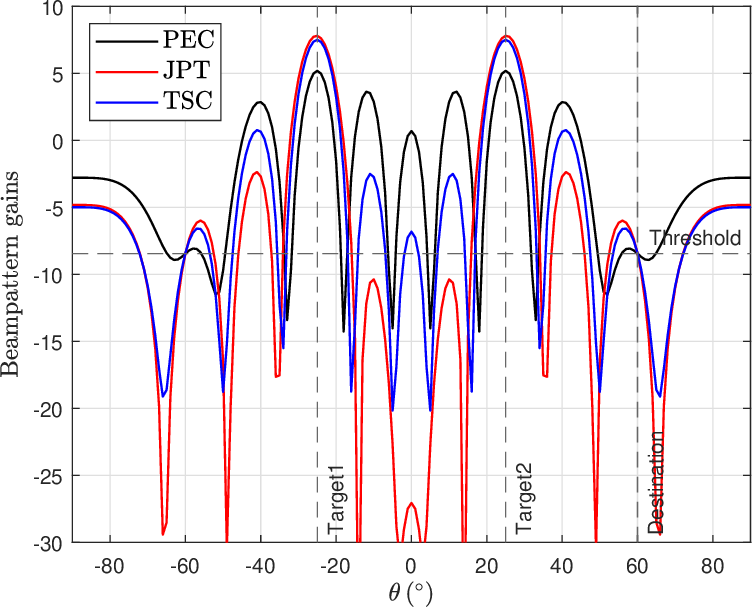}}
	\subfigure[DCRB with varying $\Delta \theta$.]{
		\label{fig04b}
		\includegraphics[width = 0.3  \textwidth]{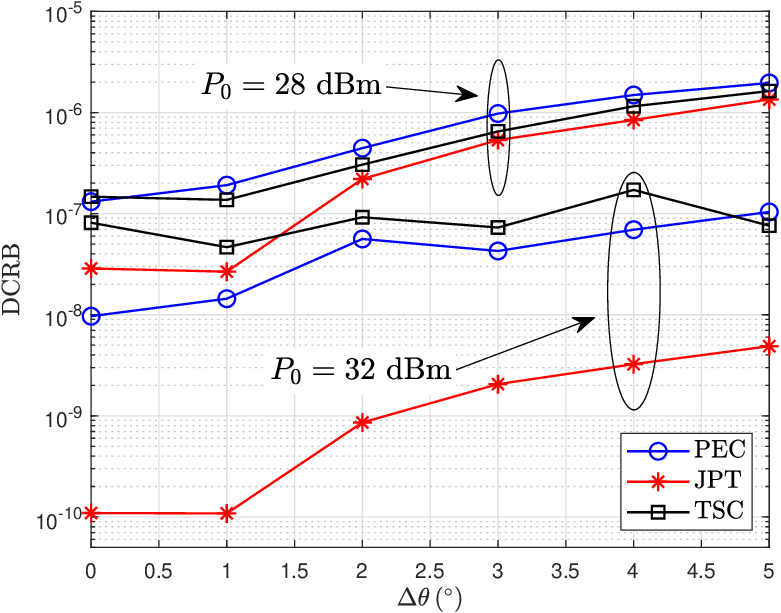}}
	\subfigure[ER for varying $\Delta \theta$.]{
		\label{fig04c}
		\includegraphics[width = 0.3  \textwidth]{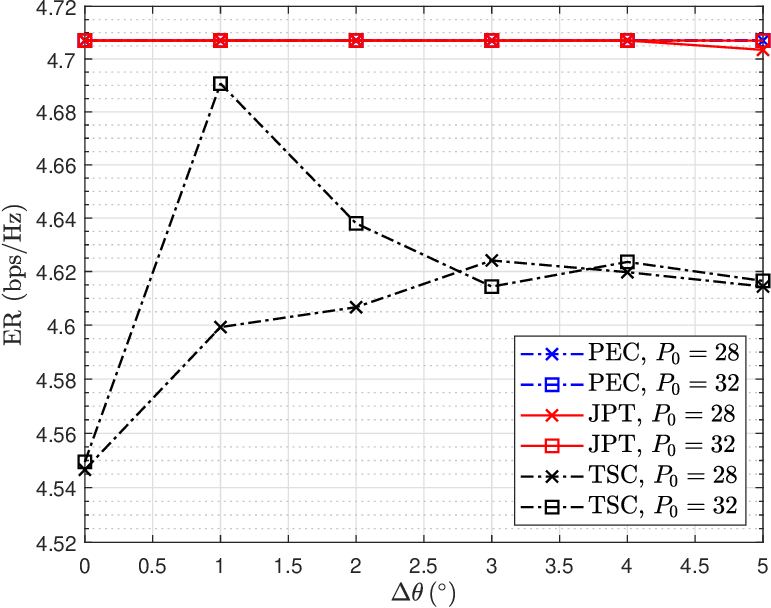}}
	\caption{{The results versus varying $P_0$ with  $\rho = 0.5$ and $N_t = 12$.}}
	\label{fig04}
\end{figure*}

Fig. \ref{fig04a} plots the beampattern gains for scenarios in which the target angles are perfectly known. 
It can be observed that the gain is maximum in the target direction and equal to the threshold in $D$'s direction. 
For the TSC scheme, the interference at $D$'s direction is slightly larger because the minimum gain in $D$'s direction is given in (\ref{interferenceCons}).
The essence of JPT design lies in the directional redistribution of beam energy by optimizing the gain of the target-direction beam while constraining energy allocation in specific directions (e.g., $D$-direction). This mechanism directly leads to its differentiated performance in non-target directions compared to TSC and PEC schemes, further validating the effectiveness of resource-coupled optimization in ISAC systems. 
Fig. \ref{fig04b} demonstrates the CRB versus varying $\Delta \theta$ under different schemes.
As the angle uncertainty increases, the sensing performance of all schemes deteriorates because a larger area must be covered with the given power.
Moreover, the change in the uncertain angle has a greater effect on JPT's sensing performance than the TSC scheme.
The sensing performance of the JPT scheme is better than that of the TSC at scenarios with higher power, for the same reasons as Figs. \ref{fig02b} and \ref{fig03b}.
Fig. \ref{fig04c} shows the ER as a function of $\Delta \theta$ for different schemes.
We find that both the PEC and JPT schemes achieve the maximum ER, independent of the uncertain angle. 
For the TSC scheme, the ER increases with the increase of the uncertain angle because the range to be covered increases, and the interference in the $D$'s direction decreases. 
Similar to Fig. \ref{fig03c}, only successfully PE is considered in the TSC scheme and the ER varies randomly.

\begin{figure*}[t]
	\centering
	\subfigure[The beampattern gains of JPT with $P_0=28$ dBm.]{
		\label{fig5a}
		\includegraphics[width = 0.21 \textwidth]{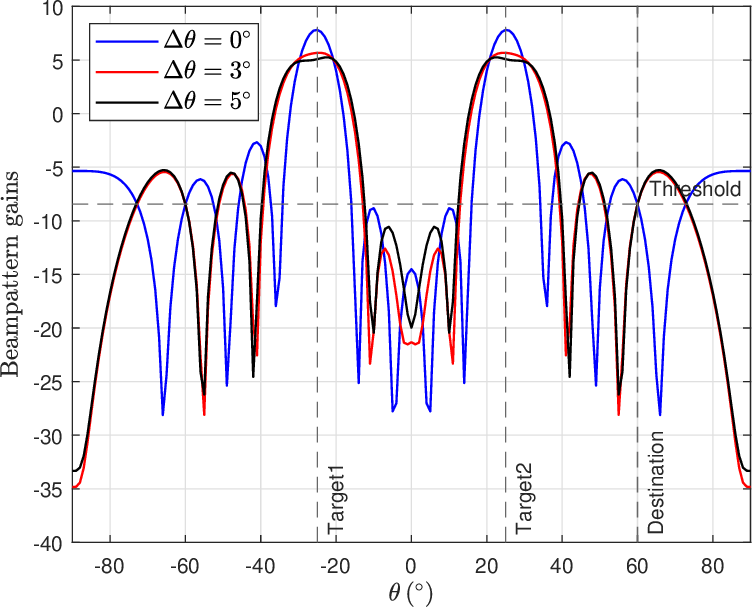}}
	\subfigure[DCRB with $P_0=28$ dBm.]{
		\label{fig5b}
		\includegraphics[width = 0.21 \textwidth]{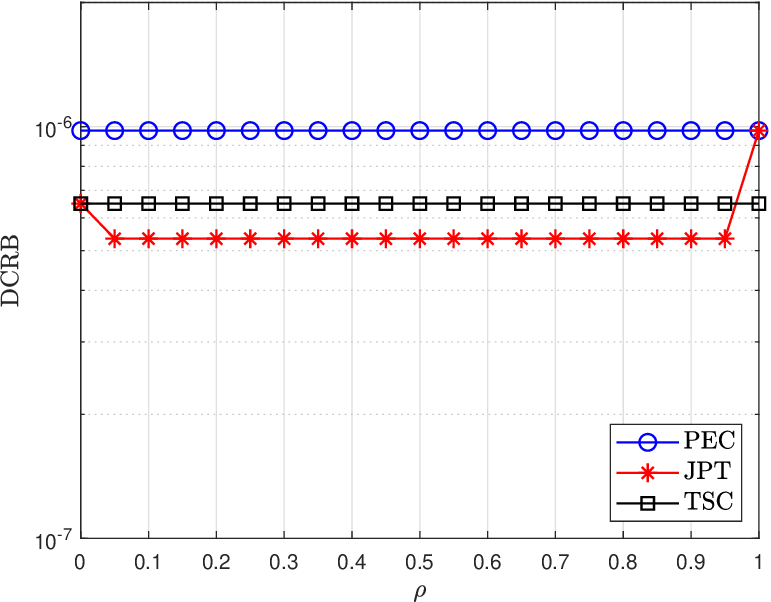}}
	\subfigure[DCRB with $P_0=30$ dBm.]{
		\label{fig5c}
		\includegraphics[width = 0.21 \textwidth]{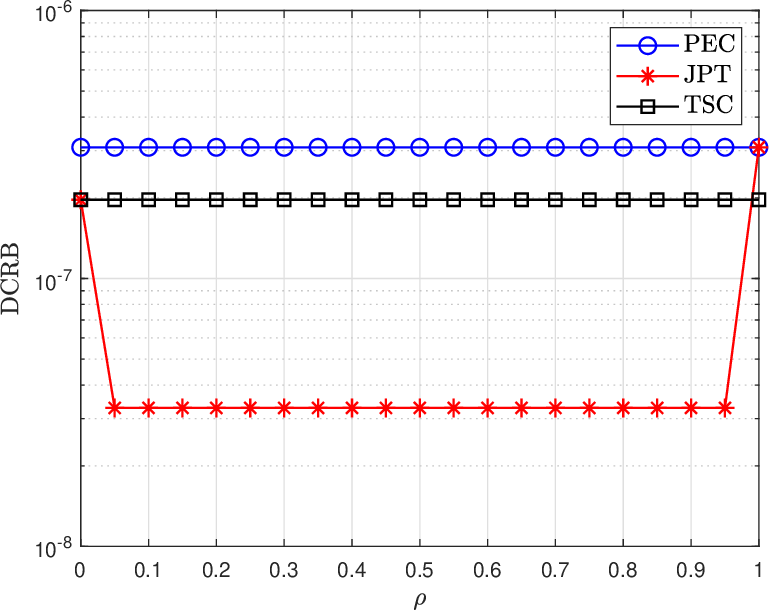}}
	\subfigure[DCRB with $P_0=32$ dBm.]{
		\label{fig5d}
		\includegraphics[width = 0.21 \textwidth]{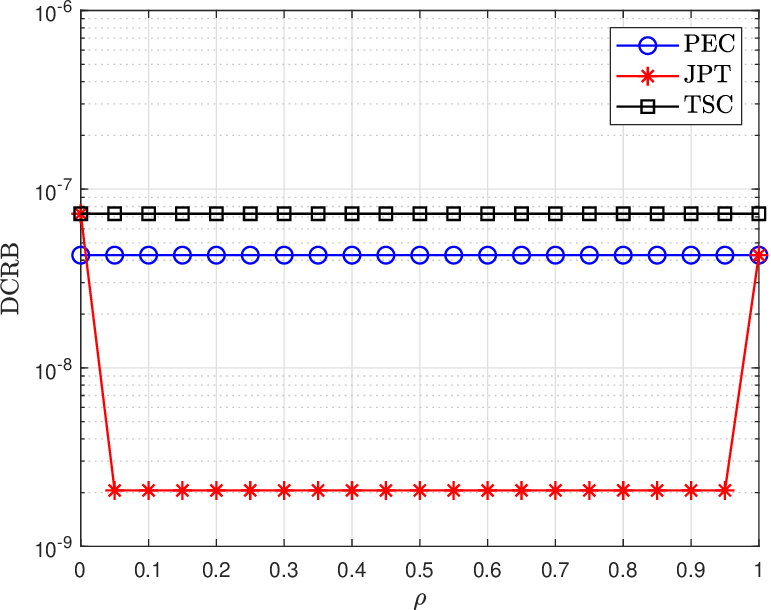}}
	\caption{The sensing performance with varying $\rho$.}
	\label{fig5}
\end{figure*}

Fig. \ref{fig5a} shows the beampattern gains of the JPT scheme with varying $\rho$. 
When $\rho = 1$, the mainlobe gain is minimum because the optimization objective in the JPT scheme reduces to minimizing the gains at $D$'s direction, which only maximizes the ER. 		
Figs. \ref{fig5b}-\ref{fig5d} plot the DCRB with varying $\rho$, respectively. 
One interesting conclusion can be observed that, in the JPT scheme, the CRB remains constant for all $\rho$ values except at the points where $\rho = 0$ and $\rho = 1$.
The reason is that when $\rho = 0$, the JPT scheme degenerates to the TSC scheme; when $\rho = 1$, it degenerates to the PEC scheme.
Comparing Figs. \ref{fig5b}-\ref{fig5d}, one can find that the sensing performance for all the schemes is improved as $P_0$ increases, which can also be found in Fig. \ref{fig02b}. 
Furthermore, the JPT's sensing performance with $\rho$ outperforms that of the TSC. 
This is because, in the JPT scheme, the beampatterns gain in $D$'s direction is kept constant to maximize the ER. Then, more power is allocated to the mainlobe, resulting in better sensing performance.

\begin{figure*}[t]
	\centering
	\subfigure[The ER with $P_0 = 28$ dBm.]{
		\label{fig6a}
		\includegraphics[width = 0.3  \textwidth]{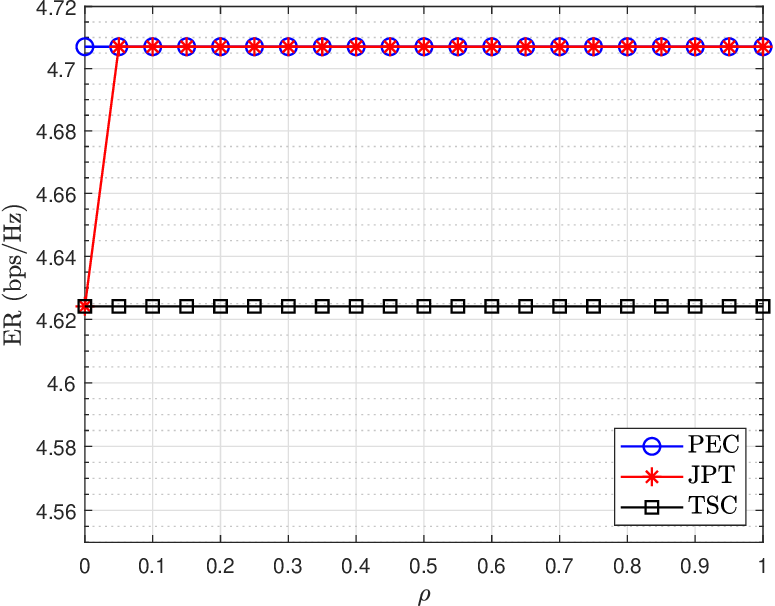}}
	\subfigure[The ER with $P_0 = 30$ dBm.]{
		\label{fig6b}
		\includegraphics[width = 0.3  \textwidth]{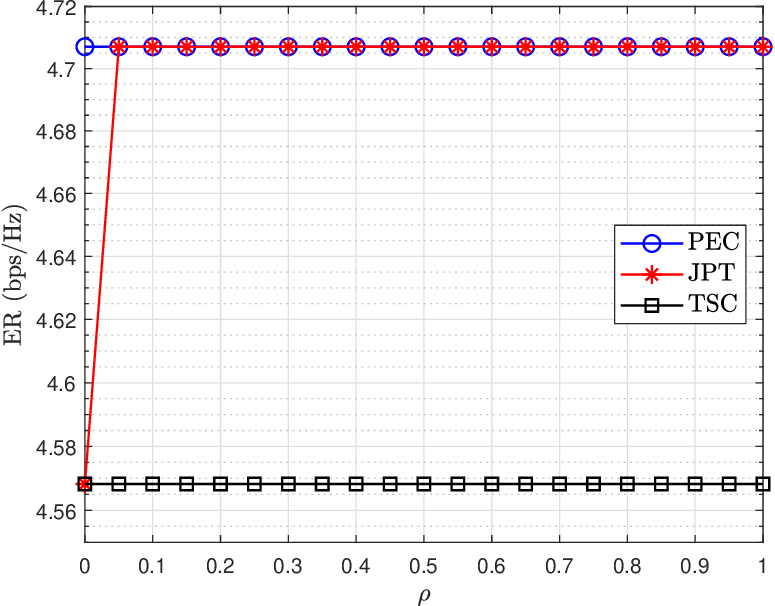}}
	\subfigure[The ER with $P_0 = 32$ dBm.]{
		\label{fig6c}
		\includegraphics[width = 0.3  \textwidth]{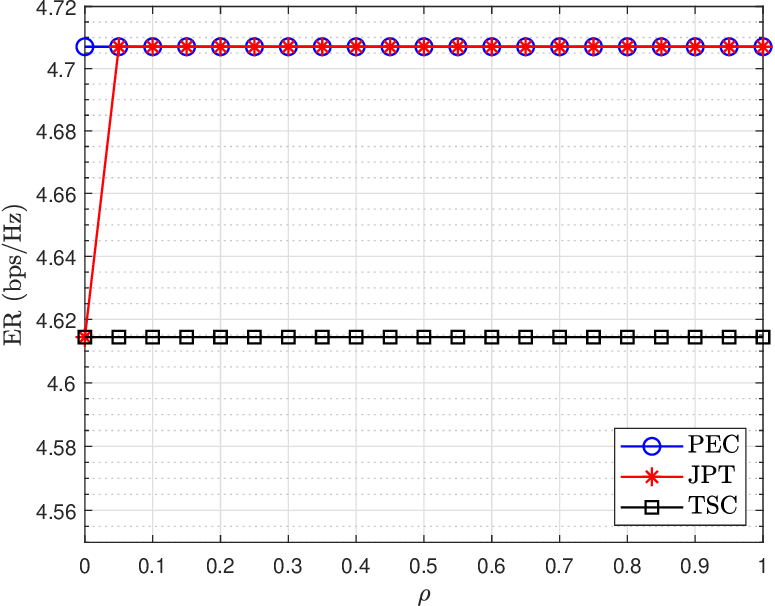}}
	\caption{The PE performance with varying $\rho$.}
	\label{fig6}
\end{figure*}

Figs. \ref{fig6a}-\ref{fig6c} plot the ER with varying $\rho$, respectively. 
We can observe that, when $\rho \ne 0$, the ER of JPT is equal to that of PEC.
For TSC, the ER exhibits a relatively weak dependence on the $P_0$, which also can be found in Fig. \ref{fig02c}. 
{{
		It must be noted that the results in Figs. \ref{fig5} and \ref{fig6} clearly demonstrate that there is no trade-off between PE and TS. 
		The core reason is as follows: the proposed joint optimization framework multiplexes radar waveforms to simultaneously transmit sensing signals and AN, which not only enables TS but also serves as AN to achieve successful PE. Specifically, by constraining the beam gain in the $D$'s direction, the ER can be maximized while ensuring successful PE.
		Meanwhile, the constrained energy in this direction enhances the degrees of freedom in energy distribution, thereby improving the TS performance. Consequently, the JPT design outperforms that of TSC and PEC.}

\begin{figure*}[t]
	\centering
	\subfigure[Beampattern gains with non-zero interference (${d}_{SE} = 1000$ m).]{
		\label{fig7a}
		\includegraphics[width = 0.3  \textwidth]{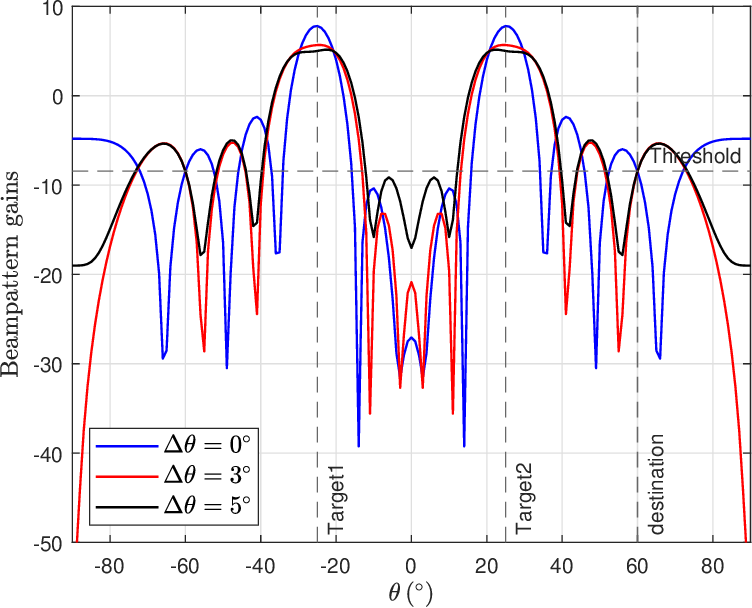}}
	\subfigure[Beampattern gains with zero interference (${d}_{SE} = 400$ m).]{
		\label{fig7b}
		\includegraphics[width = 0.3  \textwidth]{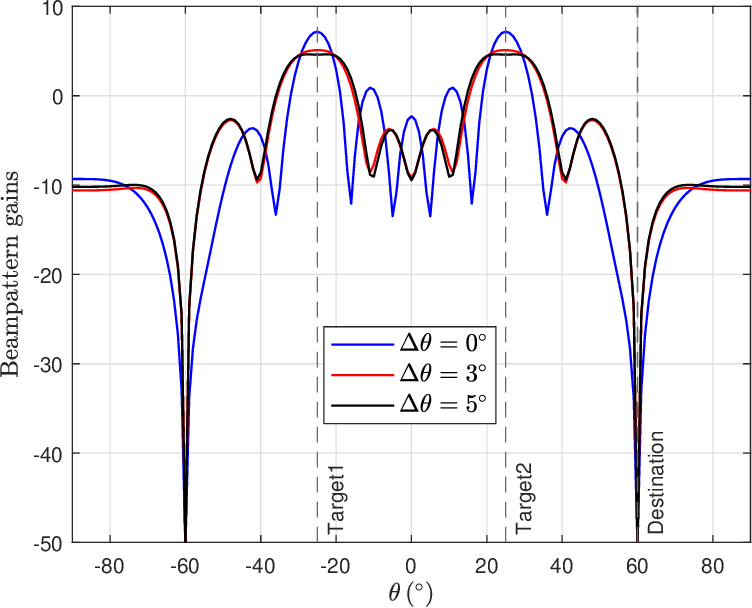}}
	\subfigure[ER for varying $\epsilon$.]{
		\label{fig7c}
		\includegraphics[width = 0.3  \textwidth]{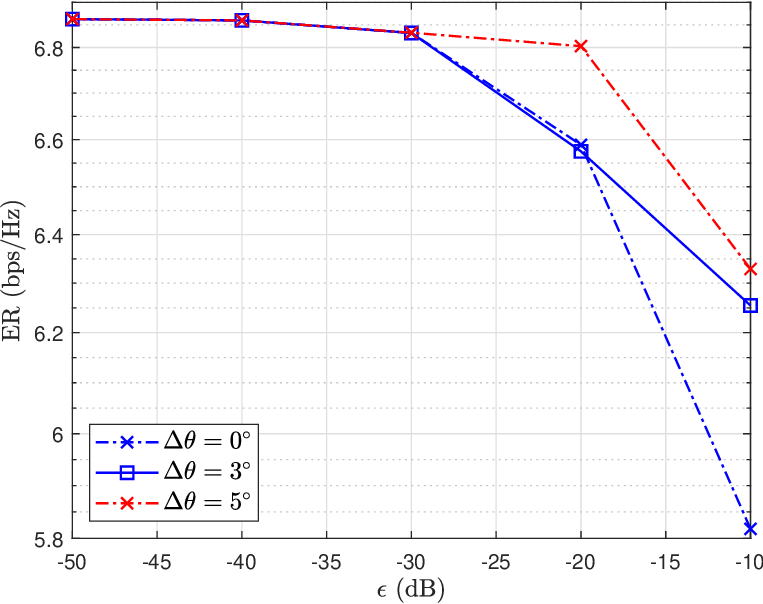}}
	\caption{The beampattern gains of the JPT scheme versus varying $\Delta \theta $.}
	\label{fig7}
\end{figure*}

Fig. \ref{fig7a} plots the beampattern gains with uncertainty angles of the JPT scheme.
We can find that, with the increase of ${\Delta \theta}$, the mainlobe width becomes more expansive and the gain decreases because the energy is spread over a larger area.
However, in the $D$'s direction, the gain must be not less than the requirement.
Fig. \ref{fig7b} plots the beampattern gains when the interference power is equal to zero. 
Because the quality of the eavesdropping channel is better than that of the illegal communication channel, it is not necessary to transmit power in $D$'s direction.
{{
		Fig. \ref{fig7c} illustrates the ER corresponding to varying $\epsilon$. 
		The results show that the ER remains constant in the lower-$\epsilon$ region.  
		As $\epsilon$ increases, the ER decreases because the interference in the $D$'s direction rises.  
}}

\begin{figure*}[t]
	\centering
	\subfigure[Beampattern gains of JPT with varying $P_0$.]{
		\label{fig08a}
		\includegraphics[width = 0.3  \textwidth]{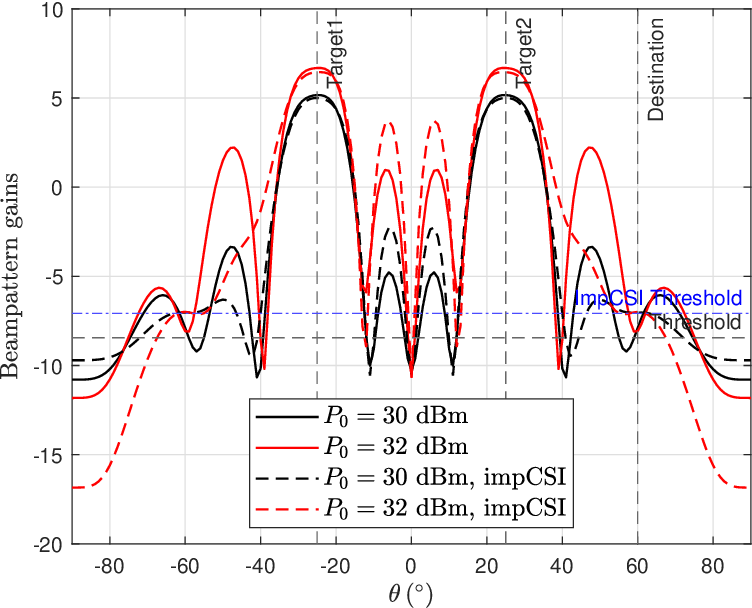}}
	\subfigure[DCRB versus varying $P_0$.]{
		\label{fig08b}
		\includegraphics[width = 0.3  \textwidth]{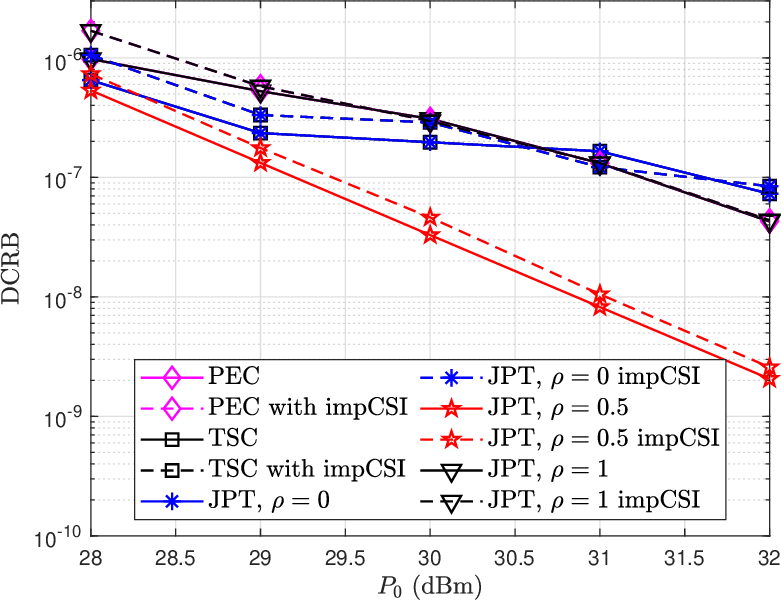}}
	\subfigure[ER versus varying $P_0$.]{
		\label{fig08c}
		\includegraphics[width = 0.3  \textwidth]{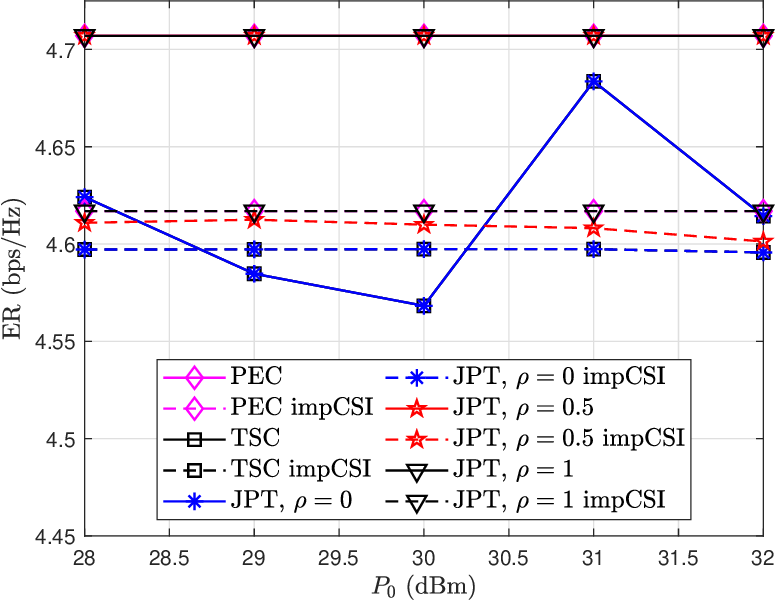}}
	\caption{The results versus varying $P_0$ with $N_t = 12$, $\Delta \theta = 3^\circ$, $\Delta \theta_D = 2^\circ$, and $\Delta {d_{ED}} = \Delta {d_{SE}} = \Delta {d_{SD}} = 30$ m.}
	\label{fig08}
\end{figure*}
{{
		To verify the robustness of the proposed algorithms, Fig. \ref{fig08} compares the results with perfect and imperfect CSI (denoted as `impCSI'). Defining $\Delta {d_{ED}}$, $\Delta {d_{SE}}$, and $\Delta {d_{SD}}$ are the distance errors, which are smaller than the exact distances. 
		According to the triangle inequality, the bound of the CSI are expressed as 
		${{\mathbf{h}}_{ED}} \ge {{{\mathbf{\hat h}}}_{ED}} = \sqrt {\frac{{{\beta _0}}}{{{{\left( {{d_{ED}} + \Delta {d_{ED}}} \right)}^\alpha }}}} {{\mathbf{a}}_t}\left( {{\hat \theta _D}} \right), {\hat \theta _D} \in \left[ {{\theta _D}- \Delta \theta _D, {\theta _D} + \Delta \theta _D } \right]$, 
		${h_{SD}} \le {{\hat h}_{SD}} = \sqrt {\frac{{{\beta _0}}}{{{{\left( {{d_{SD}} - \Delta {d_{SD}}} \right)}^\alpha }}}}$, and 
		${h_{SE}} \ge {{\hat h}_{SE}} = \sqrt {\frac{{{\beta _0}}}{{{{\left( {{d_{SE}} + \Delta {d_{SE}}} \right)}^\alpha }}}}$. 
		From Fig. \ref{fig08a}, it can be observed that under imperfect CSI, the beamgain in the target direction remains almost unchanged. In contrast, the beam in the $D$'s direction widens, demonstrating the robustness of the proposed schemes. Furthermore, the interference threshold in scenarios with imperfect CSI shows a slight increase, attributed to the increased demand for robust design considerations. Fig. \ref{fig08b} shows that imperfect CSI increases the DCRB, i.e., reduces sensing accuracy. This is because more power needs to be allocated for interference. As shown in Fig. \ref{fig08c}, imperfect CSI results in a slight decline in eavesdropping performance, as increased positional uncertainty necessitates greater power to overcome interference. Moreover, compared to the PEC scheme, the JPT scheme exhibits a more noticeable decrease in eavesdropping performance. This is because angular errors in the $D$'s direction cause the interference lobe to broaden, thereby consuming additional power resources.
		
}}

\section{Conclusion}
\label{sec:Conclusions}

This work studied the beamforming design in the joint PE and TS system wherein the multi-antenna BS transmits the signal to realize TS, which was utilized as AN to achieve PE successfully.
Firstly, two waveform design schemes were proposed to maximize the ER or minimize the CRB only.
Then, three waveform design schemes were investigated by formulating PE-centric, TS-centric, and normalized weighted optimization problems.
By omitting the rank-one constraint, the non-convex problem was solved by the SDR technique.
The SROCR method was utilized to solve continuously and iteratively to obtain the rank-one beamforming solution. 		
{{In this work, it is assumed that the CSI and location of the illegal communication links are available. However, the channel estimation errors for illicit eavesdroppers also significantly affect practical results. The scenarios wherein the BS senses the illegal transmitters and receivers, and/or with the imperfect CSI of the unlawful links, will be part of future work.}}

\begin{appendices}
	
	\section{Proof of Lemma 1 }
	\label{sec:appendicesA}	
	
	The FIM for ${\bm \zeta } = \left[ {{{\bm \theta }}, {{\bm{\beta }} _R}, {{{\bm{\beta }}}_I}} \right] $ is expressed as 
	(\ref{FIM1}), shown at the top of the next page,
	\begin{figure*}[ht]
		\begin{align}
			{\bf{F}} &= \left[ {{\bf{F}}{{\left( {\bm \zeta}  \right)}_{ij}}} \right] \nonumber\\
			&= \left[ {\frac{2}{{{\sigma ^2}}}{\rm Re tr}\left[ {\frac{{\partial {{\left( {{{\bf{A}}_r}\left( {\bm{\theta }} \right){\bm{\beta }} {\bf{A}}_t^H\left( \theta  \right){{\bf{x}}_E}} \right)}^H}}}{{\partial {\zeta _i}}}\frac{{\partial \left( {{{\bf{A}}_r}\left( {\bm{\theta }} \right){\bm{\beta }} {\bf{A}}_t^H\left( {\bm{\theta }} \right){{\bf{x}}_E}} \right)}}{{\partial {\zeta _j}}}} \right]} \right] \nonumber\\
			&= \left[ {\begin{array}{*{20}{c}}
					{{\bf{F}}\left( {{\bm{\theta }},{\bm{\theta }}} \right)}&{{\bf{F}}\left( {{\bm{\theta }},{{\bm{\beta }}_R}} \right)}&{{\bf{F}}\left( {{\bm{\theta }},{{\bm{\beta }}_I}} \right)}\\
					{{\bf{F}}\left( {{{\bm{\beta }}_R},{\bm{\theta }}} \right)}&{{\bf{F}}\left( {{{\bm{\beta }}_R},{{\bm{\beta }}_R}} \right)}&{{\bf{F}}\left( {{{\bm{\beta }}_R},{{\bm{\beta }}_I}} \right)}\\
					{{\bf{F}}\left( {{{\bm{\beta }}_I},{\bm{\theta }}} \right)}&{{\bf{F}}\left( {{{\bm{\beta }}_I},{{\bm{\beta }}_R}} \right)}&{{\bf{F}}\left( {{{\bm{\beta }}_I},{{\bm{\beta }}_I}} \right)}
			\end{array}} \right]
			\label{FIM1}
		\end{align}
		\hrulefill
	\end{figure*}
	wherein each of the elements is a $K*K$ matrix. 
	In particular, the first element is expressed as 
	\begin{align}
		{\bf{F}}\left( {{\bm{\theta }},{\bm{\theta }}} \right) = \left[ {\begin{array}{*{20}{c}}
				{F\left( {{\theta _1},{\theta _1}} \right)}&{...}&{F\left( {{\theta _1},{\theta _K}} \right)}\\
				{...}&{F\left( {{\theta _i},{\theta _j}} \right)}&{...}\\
				{F\left( {{\theta _K},{\theta _1}} \right)}&{...}&{F\left( {{\theta _K},{\theta _K}} \right)}
		\end{array}} \right],
	\end{align}		
	where $F\left( {{\theta _i},{\theta _j}} \right)$ is expressed as
	(\ref{eq26}), shown at the top of the next page, 
	\begin{figure*}[ht]
		\begin{align}
			F\left( {{\theta _i},{\theta _j}} \right) &= \frac{2}{{{\sigma ^2}}}{\mathop{\rm Re}\nolimits} {\mathop{\rm tr}\nolimits} \left[ {\frac{{\partial {{\left( {{{\bf{A}}_r}{\bm{\beta A}}_t^T{{\bf{x}}_E}} \right)}^H}}}{{\partial {\theta _i}}}\frac{{\partial \left( {{{\bf{A}}_r}{\bm{\beta A}}_t^T{{\bf{x}}_E}} \right)}}{{\partial {\theta _j}}}} \right] \nonumber \\
			&\mathop  = \limits^{\left( a \right)}  \frac{2}{{{\sigma ^2}}}{\mathop{\rm Re}\nolimits} {\mathop{\rm tr}\nolimits} \left[ \left( {{\bf{x}}_E^H{e_i}e_i^T{\dot{\bf{ A}}}_t^*{{\bm{\beta}} ^H}{\bf{A}}_r^H + {\bf{x}}_E^H{\bf{A}}_t^*{{\bm{\beta}} ^H}{e_i}e_i^T{\dot{\bf{ A}}}_r^H} \right)  \times  \left( {{{{\dot{\bf{ A}}}}_r}{e_j}e_j^T{\bm{\beta}} {\bf{A}}_t^T{{\bf{x}}_E} + {{\bf{A}}_r}{\bm{\beta}} {\dot{\bf{ A}}}_t^T{e_j}e_j^T{{\bf{x}}_E}} \right) \right] \nonumber\\
			&= \frac{2}{{{\sigma ^2}}}{\mathop{\rm Re}\nolimits} {\mathop{\rm tr}\nolimits} \left[ {{\bf{x}}_E^H{e_i}e_i^T{\bf{\dot A}}_t^*{{\bm{\beta}} ^H}{\bf{A}}_r^H{{{\bf{\dot A}}}_r}{e_j}e_j^T{\bm{\beta}} {\bf{A}}_t^T{{\bf{x}}_E} + {\bf{x}}_E^H{e_i}e_i^T{\bf{\dot A}}_t^*{{\bm{\beta}} ^H}{\bf{A}}_r^H{{\bf{A}}_r}{\bm{\beta}} {\bf{\dot A}}_t^T{e_j}e_j^T{{\bf{x}}_E}} \right. \nonumber\\
			&+ \left. {{\bf{x}}_E^H{\bf{A}}_t^*{{\bm{\beta}} ^H}{e_i}e_i^T{\bf{\dot A}}_r^H{{{\bf{\dot A}}}_r}{e_j}e_j^T{\bm{\beta}} {\bf{A}}_t^T{{\bf{x}}_E} + {\bf{x}}_E^H{\bf{A}}_t^*{{\bm{\beta}} ^H}{e_i}e_i^T{\bf{\dot A}}_r^H{{\bf{A}}_r}{\bm{\beta}} {\bf{\dot A}}_t^T{e_j}e_j^T{{\bf{x}}_E}} \right] \nonumber\\
			&\mathop  = \limits^{\left( b \right)}  \frac{2}{{{\sigma ^2}}}{\mathop{\rm Re}\nolimits} \left[ {e_i^T\left( {{\bf{A}}_r^H{{{\bf{\dot A}}}_r}} \right){e_j}e_j^T\left( {{\bm{\beta}} {\bf{A}}_t^T{{\bf{x}}_E}{\bf{x}}_E^H{\bf{\dot A}}_t^*{{\bm{\beta}} ^H}} \right){e_i} + e_i^T\left( {{\bf{A}}_r^H{{\bf{A}}_r}} \right){e_j}e_j^T\left( {{\bm{\beta}} {\bf{\dot A}}_t^T{{\bf{x}}_E}{\bf{x}}_E^H{\bf{\dot A}}_t^*{{\bm{\beta}} ^H}} \right){e_i}} \right. \nonumber\\
			& \left. { + e_i^T\left( {{\bf{\dot A}}_r^H{{{\bf{\dot A}}}_r}} \right){e_j}e_j^T\left( {{\bm{\beta}} {\bf{A}}_t^T{{\bf{x}}_E}{\bf{x}}_E^H{\bf{A}}_t^*{{\bm{\beta}} ^H}} \right){e_i} + e_i^T\left( {{\bf{\dot A}}_r^H{{\bf{A}}_r}} \right){e_j}e_j^T\left( {{\bm{\beta}} {\bf{\dot A}}_t^T{{\bf{x}}_E}{\bf{x}}_E^H{\bf{A}}_t^*{{\bm{\beta}} ^H}} \right){e_i}} \right] \nonumber\\
			&\mathop  = \limits^{\left( c \right)}  \frac{2}{{{\sigma ^2}}}{\mathop{\rm Re}\nolimits} \left[ {L{{\left( {{\bf{A}}_r^H{{{\bf{\dot A}}}_r}} \right)}_{ij}}{{\left( {{{\bm{\beta}} ^*}{\bf{\dot A}}_t^H{{\bf{R}}^*}{{\bf{A}}_t}{\bm{\beta}} } \right)}_{ij}} + L{{\left( {{\bf{A}}_r^H{{\bf{A}}_r}} \right)}_{ij}}{{\left( {{{\bm{\beta}} ^*}{\bf{\dot A}}_t^H{{\bf{R}}^*}{{{\bf{\dot A}}}_t}{\bm{\beta}} } \right)}_{ij}}} \right. \nonumber\\
			&\left. { + L{{\left( {{\bf{\dot A}}_r^H{{{\bf{\dot A}}}_r}} \right)}_{ij}}{{\left( {{{\bm{\beta}} ^*}{\bf{A}}_t^H{{\bf{R}}^*}{{\bf{A}}_t}{\bm{\beta}} } \right)}_{ij}} + L{{\left( {{\bf{\dot A}}_r^H{{\bf{A}}_r}} \right)}_{ij}}{{\left( {{{\bm{\beta}} ^*}{\bf{A}}_t^H{{\bf{R}}^*}{{{\bf{\dot A}}}_t}{\bm{\beta}} } \right)}_{ij}}} \right] \nonumber\\
			&= \frac{2}{{{\sigma ^2}}}{\left[ {{\mathop{\rm Re}\nolimits} \left( {{{\bf{F}}_{11}}} \right)} \right]_{ij}}
			\label{eq26}
		\end{align}
		\hrulefill
	\end{figure*}
	where 
	${F_{11}}  =\left( {{\bf{A}}_r^H{{\dot{\bf{ A}}}_r}} \right) \odot \left( {{{\bm{\beta}} ^*}\dot{\bf{ A}}_t^H{{\bf{R}}^*}{{\bf{A}}_t}{\bm{\beta}} } \right) + \left( {{\bf{A}}_r^H{{\bf{A}}_r}} \right) \odot \left( {{{\bm{\beta}} ^*}\dot{\bf{ A}}_t^H{{\bf{R}}^*}{{\dot{\bf{ A}}}_t}{\bm{\beta}} } \right) + \left( {\dot{\bf{ A}}_r^H{{\dot{\bf{ A}}}_r}} \right) \odot \left( {{{\bm{\beta}} ^*}{\bf{A}}_t^H{{\bf{R}}^*}{{\bf{A}}_t}{\bm{\beta}} } \right) + \left( {\dot{\bf{ A}}_r^H{{\bf{A}}_r}} \right) \odot \left( {{{\bm{\beta}} ^*}{\bf{A}}_t^H{{\bf{R}}^*}{{\dot{\bf{ A}}}_t}{\bm{\beta}} } \right)$, 
	${\left[ {\bf{{ B}}} \right]_{ij}}$ denotes the $(i,j)$ element of the matrix $\bf{{ B}}$, 
	$e_i$ denotes the $i$-th column of the identity matrix, 
	step $\left( a \right)$ holds by following 
	\begin{align}
		\frac{{\partial \left( {{{\bf{A}}_r}{\bm{\beta}} {\bf{A}}_t^T{{\bf{x}}_E}} \right)}}{{\partial {\theta _i}}} &= {{\dot{\bf{ A}}}_r}{e_i}e_i^T{\bm{\beta}} {\bf{A}}_t^T{{\bf{x}}_E}  + {{\bf{A}}_r}{\bm{\beta}} {\dot{\bf{ A}}}_t^T{e_i}e_i^T{{\bf{x}}_E},
	\end{align}
	step $\left( b \right)$ holds by following 
	\begin{align}
		&{\mathop{\rm tr}\nolimits} \left[{\left( {{\bf{x}}_E^H{e_i}e_i^T{\dot{\bf{ A}}}_t^*{{\bm{\beta}} ^H}{\bf{A}}_r^H} \right)\left( {{{{\dot{\bf{ A}}}}_r}{e_j}e_j^T{\bm{\beta}} {\bf{A}}_t^T{{\bf{x}}_E}} \right)}\right]  \nonumber\\
		&= e_i^T\left( {{\bf{A}}_r^H{{{\dot{\bf{ A}}}}_r}} \right){e_j}e_j^T\left( {{\bm{\beta}} {\bf{A}}_t^T{{\bf{x}}_E}{\bf{x}}_E^H{\dot{\bf{ A}}}_t^*{{\bm{\beta}} ^H}} \right){e_i} \nonumber\\
		&= L{\left( {{\bf{A}}_r^H{{{\dot{\bf{ A}}}}_r}} \right)_{ij}}{\left( {{{\bm{\beta}} ^*}{\dot{\bf{ A}}}_t^H{{\bf{R}}^*}{{\bf{A}}_t}{\bm{\beta}} } \right)_{ij}},
	\end{align} 
	and
	step $\left( c \right)$ holds by following 
	\begin{align}
		&L{\left( {{\bf{A}}_r^H{{{\dot{\bf{ A}}}}_r}} \right)_{ij}}{\left( {{{\bm{\beta}} ^*}{\dot{\bf{ A}}}_t^H{{\bf{R}}^*}{{\bf{A}}_t}{\bm{\beta}} } \right)_{ij}} \nonumber\\
		&= {\left[ {{\bf{A}}_r^H{{{\dot{\bf{ A}}}}_r} \odot {{\bm{\beta}} ^*}{\dot{\bf{ A}}}_t^H{{\bf{R}}^*}{{\bf{A}}_t}{\bm{\beta}} } \right]_{ij}}.
	\end{align}

	With some simple algebraic manipulations, the element of ${\bf{F}}\left( {{\bm{\theta }},{{\bm{{\bm{\beta}} }}_R}} \right)$ is expressed as
	(\ref{eq30}), shown at the top of the next page, 
	\begin{figure*}[ht]
		\begin{align}
			F\left( {{\theta _i},{b_{{R_j}}}} \right) &= \frac{2}{{{\sigma ^2}}}{\mathop{\rm Re}\nolimits} {\mathop{\rm tr}\nolimits}  \left[{\frac{{\partial {{\left( {{{\bf{A}}_r}{\bm{\beta}} {\bf{A}}_t^T{{\bf{x}}_E}} \right)}^H}}}{{\partial {\theta _i}}}\frac{{\partial \left( {{{\bf{A}}_r}{\bm{\beta}} {\bf{A}}_t^T{{\bf{x}}_E}} \right)}}{{\partial {b_{{R_j}}}}}}\right]  \nonumber\\
			&= \frac{2}{{{\sigma ^2}}}{\mathop{\rm Re}\nolimits}  \left[L{{\left( {{\bf{A}}_r^H{{\bf{A}}_r}} \right)}_{ij}}{{\left( {{{\bm{\beta}} ^*}{\dot{\bf{ A}}}_t^H{{\bf{R}}^*}{{\bf{A}}_t}} \right)}_{ij}} \right.  \left. + L{{\left( {{\dot{\bf{ A}}}_r^H{{\bf{A}}_r}} \right)}_{ij}}{{\left( {{{\bm{\beta}} ^*}{\bf{A}}_t^H{{\bf{R}}^*}{{\bf{A}}^t}} \right)}_{ij}}\right]  \nonumber \\
			&= \frac{2}{{{\sigma ^2}}}{\left[ {{\mathop{\rm Re}\nolimits} \left( {{{\bf{F}}_{12}}} \right)} \right]_{ij}}
			\label{eq30}
		\end{align}
		\hrulefill
	\end{figure*}
	where 
	${F_{12}} = \left( {{\bf{A}}_r^H{{\bf{A}}_r}} \right) \odot \left( {{{\bm{\beta}} ^*}\dot{\bf{ A}}_t^H{{\bf{R}}^*}{{\bf{A}}_t}} \right) + \left( {\dot{\bf{ A}}_r^H{{\bf{A}}_r}} \right) \odot \left( {{{\bm{\beta}} ^*}{\bf{A}}_t^H{{\bf{R}}^*}{{\bf{A}}_t}} \right)$ 
	and 
	$\frac{{\partial \left( {{{\bf{A}}_r}{\bm{\beta}} {\bf{A}}_t^T{{\bf{x}}_E}} \right)}}{{\partial {b_{{R_j}}}}} = {{\bf{A}}_r}{e_j}e_j^T{\bf{A}}_t^T{{\bf{x}}_E}$. 
	
	The element of ${\bf{F}}\left( {{\bm{\theta }},{{\bm{{\bm{\beta}} }}_I}} \right)$, 
	${{\bf{F}}\left( {{{\bm{{\bm{\beta}} }}_R},{\bm{\theta }}} \right)}$, 
	${{\bf{F}}\left( {{{\bm{{\bm{\beta}} }}_R},{{\bm{{\bm{\beta}} }}_R}} \right)}$, 
	${{\bf{F}}\left( {{{\bm{{\bm{\beta}} }}_R},{{\bm{{\bm{\beta}} }}_I}} \right)}$, 
	${{\bf{F}}\left( {{{\bm{{\bm{\beta}} }}_I},{\bm{\theta }}} \right)}$, 
	${{\bf{F}}\left( {{{\bm{{\bm{\beta}} }}_I},{{\bm{{\bm{\beta}} }}_R}} \right)}$, and 
	${{\bf{F}}\left( {{{\bm{{\bm{\beta}} }}_I},{{\bm{{\bm{\beta}} }}_I}} \right)}$ are expressed as
	(\ref{eq31})-(\ref{eq37}), shown at the top of the last page, where 
	${F_{22}} = \left( {{\bf{A}}_r^H{{\bf{A}}_r}} \right) \odot \left( {{\bf{A}}_t^H{{\bf{R}}^*}{{\bf{A}}_t}} \right)$
	and 
	$\frac{{\partial \left( {{{\bf{A}}_r}{\bm{\beta}} {\bf{A}}_t^T{{\bf{x}}_E}} \right)}}{{\partial {b_{{I_j}}}}} = j{{\bf{A}}_r}{e_j}e_j^T{\bf{A}}_t^T{{\bf{x}}_E}$.
	
	\begin{figure*}[t]
		\begin{align}
			F\left( {{\theta _i},{b_{{I_j}}}} \right) &= \frac{2}{{{\sigma ^2}}}{\mathop{\rm Re}\nolimits} {\mathop{\rm tr}\nolimits}  \left[{\frac{{\partial {{\left( {{{\bf{A}}_r}{\bm{\beta}} {\bf{A}}_t^T{{\bf{x}}_E}} \right)}^H}}}{{\partial {\theta _i}}}\frac{{\partial \left( {{{\bf{A}}_r}{\bm{\beta}} {\bf{A}}_t^T{{\bf{x}}_E}} \right)}}{{\partial {b_{{I_j}}}}}}\right]  \nonumber\\
			&= \frac{2}{{{\sigma ^2}}}{\mathop{\rm Re}\nolimits}  \left[L{{\left( {j{\bf{A}}_r^H{{\bf{A}}_r}} \right)}_{ij}}{{\left( {{{\bm{\beta}} ^*}{\dot{\bf{ A}}}_t^H{{\bf{R}}^*}{{\bf{A}}_t}} \right)}_{ij}} \right. \left. + L{{\left( {j{\dot{\bf{ A}}}_r^H{{\bf{A}}_r}} \right)}_{ij}}{{\left( {{{\bm{\beta}} ^*}{\bf{A}}_t^H{{\bf{R}}^*}{{\bf{A}}^t}} \right)}_{ij}} \right] \nonumber\\
			&=  - \frac{2}{{{\sigma ^2}}}{\left[ {{\mathop{\rm Im}\nolimits} \left( {{{\bf{F}}_{12}}} \right)} \right]_{ij}}
			\label{eq31}
		\end{align}
		\hrulefill
		\begin{align}
			F\left( {{b_{{R_i}}},{\theta _j}} \right) 
			&= \frac{2}{{{\sigma ^2}}}{\mathop{\rm Re}\nolimits}  \left[L{{\left( {{\bf{A}}_r^H{{\bf{A}}_r}} \right)}_{ij}}{{\left( {{\bf{A}}_t^H{{\bf{R}}^*}{{{\dot{\bf{ A}}}}_t}{\bm{\beta}} } \right)}_{ij}} \right.  \left. +L{{\left( {{\bf{A}}_r^H{{{\dot{\bf{ A}}}}_r}} \right)}_{ij}}{{\left( {{\bf{A}}_t^H{{\bf{R}}^*}{{\bf{A}}_t}{\bm{\beta}} } \right)}_{ij}}\right] \nonumber\\
			&= \frac{2}{{{\sigma ^2}}}{\left[ {{{{\mathop{\rm Re}\nolimits} }^T}\left( {{{\bf{F}}_{12}}} \right)} \right]_{ij}}
			\label{eq32}
		\end{align}
		\hrulefill
		\begin{align}
			F\left( {{b_{{R_i}}},{b_{{R_j}}}} \right) &= \frac{2}{{{\sigma ^2}}}{\mathop{\rm Re}\nolimits} {\mathop{\rm tr}\nolimits}  \left[{\left( {{\bf{x}}_E^H{\bf{A}}_t^*{e_i}e_i^T{\bf{A}}_r^H} \right)\left( {{{\bf{A}}_r}{e_j}e_j^T{\bf{A}}_t^T{{\bf{x}}_E}} \right)}\right]  \nonumber \\
			&= \frac{2}{{{\sigma ^2}}}{\mathop{\rm Re}}  \left[{L{{\left( {{\bf{A}}_r^H{{\bf{A}}_r}} \right)}_{ij}}{{\left( {{\bf{A}}_t^H{{\bf{R}}^*}{{\bf{A}}_t}} \right)}_{ij}}}\right] \nonumber\\
			&= \frac{2}{{{\sigma ^2}}}{\left[ {{\mathop{\rm Re}} \left( {{{\bf{F}}_{22}}} \right)} \right]_{ij}}
			\label{eq33}
		\end{align}
		\hrulefill
		\begin{align}
			F\left( {{b_{{R_i}}},{b_{{I_j}}}} \right) &= \frac{2}{{{\sigma ^2}}}{\mathop{\rm Re}} {\mathop{\rm tr}} \left[{\left( {{\bf{x}}_E^H{e_i}e_i^T{\bf{A}}_t^*{\bf{A}}_r^H} \right)\left( {j{{\bf{A}}_r}{e_j}e_j^T{\bf{A}}_t^T{{\bf{x}}_E}} \right)}\right] \nonumber\\
			&= \frac{2}{{{\sigma ^2}}}{\mathop{\rm Re}}  \left[{L{{\left( {j{\bf{A}}_r^H{{\bf{A}}_r}} \right)}_{ij}}{{\left( {{\bf{A}}_t^H{{\bf{R}}^*}{{\bf{A}}_t}} \right)}_{ij}}}\right] \nonumber\\
			&=  - \frac{2}{{{\sigma ^2}}}{\left[ {{{{\mathop{\rm Im}} }^T}\left( {{{\bf{F}}_{22}}} \right)} \right]_{ij}}
			\label{eq34}
		\end{align}
		\hrulefill
		\begin{align}
			F\left( {{b_{{I_i}}},{\theta _j}} \right) &= \frac{2}{{{\sigma ^2}}}{\mathop{\rm Re}\nolimits}  \left[L{{\left( { - j{\bf{A}}_r^H{{{\dot{\bf{ A}}}}_r}} \right)}_{ij}}{{\left( {{\bf{A}}_t^H{{\bf{R}}^*}{{\bf{A}}_t}{\bm{\beta}} } \right)}_{ij}} \right. \left.+L{{\left( { - j{\bf{A}}_r^H{{{\dot{\bf{ A}}}}_r}} \right)}_{ij}}{{\left( {{\bf{A}}_t^H{{\bf{R}}^*}{{\bf{A}}_t}{\bm{\beta}} } \right)}_{ij}} \right]  \nonumber\\
			&=  - \frac{2}{{{\sigma ^2}}}{\left[ {{{{\mathop{\rm Im}\nolimits} }^T}\left( {{{\bf{F}}_{12}}} \right)} \right]_{ij}}
			\label{eq35}
		\end{align}
		\hrulefill
		\begin{align}
			F\left( {{b_{{I_i}}},{b_{{R_j}}}} \right) &= \frac{2}{{{\sigma ^2}}}{\mathop{\rm Re}\nolimits} {\rm tr}  \left[{\left( { - j{\bf{x}}_E^H{\bf{A}}_t^*{e_i}e_i^T{\bf{A}}_r^H} \right)\left( {{{\bf{A}}_r}{e_j}e_j^T{\bf{A}}_t^T{{\bf{x}}_E}} \right)}\right]  \nonumber\\
			&= \frac{2}{{{\sigma ^2}}}{\mathop{\rm Re}\nolimits} \left[{{{L\left( { - j{\bf{A}}_r^H{{\bf{A}}_r}} \right)}_{ij}}{{\left( {{\bf{A}}_t^H{{\bf{R}}^*}{{\bf{A}}_t}} \right)}_{ij}}}\right]  \nonumber\\
			&=  - \frac{2}{{{\sigma ^2}}}{\left[ {{{{\mathop{\rm Im}\nolimits} }^T}\left( {{{\bf{F}}_{22}}} \right)} \right]_{ij}}
			\label{eq36}
		\end{align}
		\hrulefill
		\begin{align}
			F\left( {{b_{{I_i}}},{b_{{I_j}}}} \right) &= \frac{2}{{{\sigma ^2}}}{\mathop{\rm Re}\nolimits} {\rm tr}  \left[{\left( { - j{\bf{x}}_E^H{\bf{A}}_t^*{e_i}e_i^T{\bf{A}}_r^H} \right)\left( {j{{\bf{A}}_r}{e_j}e_j^T{\bf{A}}_t^T{{\bf{x}}_E}} \right)}\right]   \nonumber\\
			&= \frac{2}{{{\sigma ^2}}}{\mathop{\rm Re}\nolimits}  \left[L {{{\left( {{\bf{A}}_r^H{{\bf{A}}_r}} \right)}_{ij}}{{\left( {{\bf{A}}_t^H{{\bf{R}}^*}{{\bf{A}}_t}} \right)}_{ij}}}\right] \nonumber\\
			&= \frac{2}{{{\sigma ^2}}}{\left[ {{{{\mathop{\rm Re}\nolimits} }^T}\left( {{{\bf{F}}_{22}}} \right)} \right]_{ij}}
			\label{eq37}
		\end{align}		
	\end{figure*}
	
\end{appendices}

\end{document}